\documentclass{IEEEtran}
\usepackage{cite}
\usepackage{mathtools}
\usepackage{amsmath,amssymb,amsfonts}
\usepackage{algorithmic}
\usepackage{graphicx}
\usepackage{textcomp}
\usepackage[capitalize]{cleveref}
\usepackage[dvipsnames]{xcolor}
\crefformat{equation}{(#2#1#3)}
\crefrangeformat{equation}{#3(#1)#4--#5(#2)#6}
\usepackage{supertabular}
\usepackage{multicol}
\usepackage{multirow}
\usepackage{caption}
\usepackage[normalem]{ulem}

\def\BibTeX{{\rm B\kern-.05em{\sc i\kern-.025em b}\kern-.08em
    T\kern-.1667em\lower.7ex\hbox{E}\kern-.125emX}}
\begin{document}
\bstctlcite{URLBSTcontrol}
\title{Eliminating Reflections in Waveguide Bends Using a Metagrating-Inspired Semianalytical Methodology}
\author{Liran Biniashvili, \IEEEmembership{Student Member, IEEE} and Ariel Epstein, \IEEEmembership{Senior Member, IEEE}
\thanks{This work was supported by the Israel Science Foundation (Grant 1540/18).}
\thanks{The authors are with the Andrew and Erna Viterbi Faculty of Electrical Engineering, Technion - Israel Institute of Technology, Haifa 3200003, Israel (e-mail: liranbinia@campus.technion.ac.il; epsteina@ee.technion.ac.il).}}

\maketitle
\captionsetup[figure]{font=footnotesize}

\begin{abstract}
We present a semianalytical method to obtain perfect transmission across abrupt H-plane bends in single-mode rectangular waveguides using a single passive polarizable element (scatterer). The underlying analysis and synthesis schemes are inspired by the rapidly-growing research on metagratings, typically used to manipulate wave trajectories in free-space. These sparse configurations of subwavelength polarizable particles (meta-atoms) are designed by careful tailoring of inter-element near-field and far-field interactions, relying on analytical models to resolve the required meta-atom distribution and geometry to facilitate a desired interference pattern when excited by the incident wave. Utilizing these metagrating design concepts, we develop a modal formalism for obtaining a collection of locations inside the bend junction, in which a passive scatterer may be placed to zero out the return loss. Subsequently, we propose two different shapes for the scatterer and discuss, for each of them, the ways in which their geometrical characteristics may be retrieved. This versatile and efficient methodology, verified via full-wave simulations, can be utilized to eliminate reflection loss in diverse bend configurations, often found in complex wave-guiding systems used for antenna feeding and power transmission. Moreover, these results demonstrate the usefulness and potential of  metagrating design concepts for various applications, beyond free-space beam manipulation.
\end{abstract}

\begin{IEEEkeywords}
waveguides, metagratings, mode-matching, radial waveguide, waveguide bend
\end{IEEEkeywords}

\section{Introduction}
\label{sec:introduction}
\IEEEPARstart{F}{or} many years, waveguides (WGs) have been a crucial part in a variety of radio frequency devices, used for transmission of electromagnetic power in diverse applications, from satellite communication systems to millimeter-wave components \cite{pozar2011microwave}. As a part of the design of such systems, physical constraints may often require bending of the WGs along their path, which typically results in reduced performance due to reflection loss. Specifically, the bend constitutes a sudden change in the boundary conditions, leading to undesired scattering of the guided modes at the formed junction.

Consequently, overcoming and controlling scattering in WG bends have been of major interest throughout the years. A common approach for eliminating these reflections involves optimization of the bend geometry, i.e. using - instead of the sharp abrupt corners - stepped \cite{Carle1991,Casanueva2004,Manzillo2017}, mitered \cite{Alessandri1994,Reiter1994,Casanueva2004,Ma1997,Alimenti1994} and curved \cite{Coccioli1996} bends, to name a few. In addition, right-angle WG bend compensation has been suggested by introducing some scattering mechanism into the junction, e.g. a partial-height metallic cylindrical post \cite{partialheightcompensate}, a square dielectric post (alongside bend mitering) \cite{Wu1989}, or a matching septum (diaphragm) installed in the middle of a curved bend \cite{Mongiardo1995}.

Metamaterials have also been proposed to increase the transmission efficiency in bends, either by adding a complex media section inside the WG, using concepts such as transformation optics \cite{MMs16,MMs7,MMs9}, $\epsilon$-near-zero effects \cite{MMs5}, negative-index \cite{MMs12} or geometrical optics \cite{MMs14,MMs10}, or by introducing metamaterial-inspired elements inside the junction \cite{MMs1} or in-between two given rectangular WGs \cite{MMs4}. Furthermore, in \cite{ms_in_wg} we have discussed utilizing an in-line metasurface as means to suppress WG bend reflections.

Most of these methods, however, seem to rely on dedicated fabrication of a relatively cumbersome bend structure, which may be challenging. Moreover, many of these approaches are often applicable to a limited range of configurations, e.g. only to right-angle bends and/or symmetric bends (where input and output ports are of identical widths).

In this paper, we propose a different approach for achieving seamless WG bends, relying on an efficient and insightful analytical model to devise a simple solution, based on a single scatterer, without requiring any deformation of the bend. Importantly, utilizing a general model attributes flexibility to this solution scheme, supporting a wide variety of bend angles and WG widths, with no symmetry constraints on the input and output.

The methodology presented herein is inspired by recent work on metagratings \cite{PhysRevLett.119.067404,PhysRevApplied.8.054037,Rabinovich2018,Popov2018,Rabinovich_2,Casolaro2020,Rabinovich_3}. Metagratings (MGs) are sparse periodic arrays of subwavelength polarizable elements, used for obtaining highly-efficient control over diffraction of plane waves in free-space \cite{MgGeneral1,MgGeneral2,MgGeneral3,MgGeneral4,MgGeneral6}. As opposed to metasurfaces, which consist of densely arranged meta-atoms, effectively forming homogenized electromagnetic properties \cite{GLYBOVSKI2016,Epstein2016,Caloz2017}, the engineering of MG functionality involves detailed scattering analysis, considering the secondary fields arising from the currents induced on the individual meta-atoms. Subsequently, the desired functionality is achieved by arranging the meta-atoms and choosing their geometry such that the emerging interference pattern fits the design goals. This scheme is facilitated by an analytical model deriving from the inherent periodicity of the MGs, enabling the scattered fields to be expressed as a superposition of Floquet-Bloch modes (only few of which are propagating). The target device response is then obtained by formulating constraints on the scattering parameters and using the system degrees of freedom (DOFs) to find a configuration that satisfies them. This results in a reliable semianalytical design procedure, yielding a detailed configuration blueprint, typically exhibiting high performance without the need of extensive full-wave optimization. Therefore, and since both the diffraction problem for plane waves and the scattering problem for guided modes require control of the power coupling to a finite number of propagating modes, we were motivated to explore the possibility to harness modal analysis and extend the appealing MG synthesis approach to devise a solution to the WG bend reflection loss considered herein.

In particular, we propose solving the problem of abrupt H-plane bends in single-mode rectangular WGs by placing a single polarizable element ("scatterer", or "meta-atom") inside the bending region of the WG \cite{MG_WG_BEND2019} (\cref{WG_drawing}). To this end, we extend the formalism developed by Widarta, Kuwano, et al. \cite{Widarta1994,Kuwano2003}, treating the bend in the frame of modal analysis with the boundary contour mode-matching (BCMM) method, to enable rigorous inclusion of the scatterer in the bend for calculation of the system scattering parameters. As will be shown, these scattering coefficients depend on the scatterer location in the bend and on the current induced on it. Therefore, in analogy to the MG design procedure \cite{PhysRevLett.119.067404,PhysRevApplied.8.054037}, a suitable choice of the scatterer location and dimensions can yield full transmission of power from input to output.

As in the case of free-space beam-manipulating MGs \cite{PhysRevLett.119.067404,PhysRevApplied.8.054037}, we first model the scatterer by a current line whose current and location must follow specific constraints to ensure system passivity and unitary transmission across the bend. The constraints on the scattering parameters are applied directly on the semianalytical model, allowing efficient resolution of these DOFs. Next, to enable practical realization of such a solution, we replace the abstract current with a suitable passive polarizable element, and show that its geometry can be tuned to facilitate perfect bend matching as long as it is appropriately placed within the junction.

We verify the semianalytical model and test the synthesis procedure via full-wave simulations (ANSYS HFSS). In these case studies, we demonstrate the flexibility the designer has in choosing the meta-atom configuration by using two different scatterer geometries: a capacitively-loaded wire (similarly to \cite{PhysRevApplied.8.054037}) and a cylindrical metallic post. The latter is found to be especially useful, as it further allows semianalytical estimation of the post radius required to obtain optimal transmission efficiency, enabling complete execution of the synthesis procedure, from user-defined goals to fabrication-ready design specifications, while avoiding long full-wave optimizations.

It should be noted that scatterers are widely used in WGs \cite{scatterers7,scatterers10,scatterers11,scatterers2,scatterers8,scatterers4,scatterers6}, and in particular, as mentioned above, as means to reduce reflection loss in WG bends \cite{partialheightcompensate,Wu1989,Mongiardo1995}. In contrast to these approaches for WG bend matching, which inherently include assumptions for the scatterer shape and/or location in the WG, our metagrating-inspired method - which relies on an abstract current-line model - is versatile both in terms of the scatterer shape and its location, and may allow better utilization of the design DOFs, e.g., to meet fabrication constraints or to improve bandwidth.

Importantly, as shall be discussed in \cref{subsec:branches}, relying on an analytical model provides fundamental physical insight regarding the nature of the emerging solution, eventually assisting in choosing design paths for enhanced performance. Overall, the proposed methodology establishes an efficient and flexible engineering tool for solving the WG bend problem, highlighting the potential of the MG synthesis approach to tackle challenges in a variety of electromagnetic scenarios, beyond free-space beam manipulation.

\section{Theory}
\label{sec:theory}
\subsection{System Configuration}
\label{subsec:system_configuration}
\begin{figure}[!t]
\centerline{\includegraphics[width=\columnwidth]{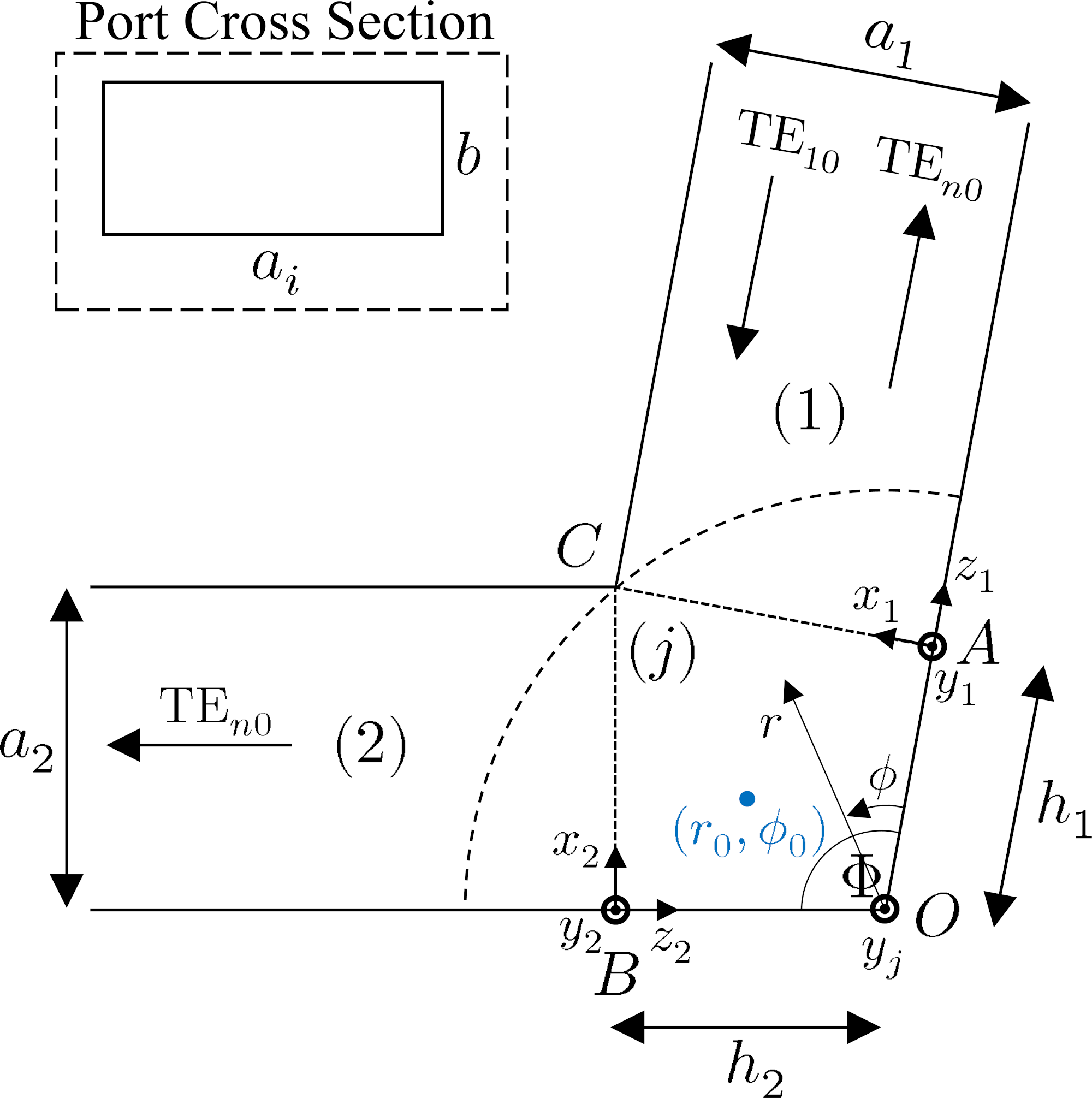}}
\caption{Physical configuration of a rectangular waveguide with a general abrupt H-plane bend (top view). The excitation propagates to the junction from port $(1)$, and a scatterer is placed at $(r_0,\phi_0)$ inside the junction $(j)$. The inset in the upper-left corner illustrates the cross section of the rectangular ports ($i\in\{1,2\}$).}
\label{WG_drawing}
\end{figure}

We consider a single-mode WG with perfectly conducting walls, bent at an arbitrary angle $\Phi$, as shown in \cref{WG_drawing}. The WG is divided into three regions: region $(1)$ includes the straight WG section stretching from the input port to the bend, region $(2)$ extends from the bend to the output port, and the junction area is referred to as region $(j)$. The height along the WG is $b$, and in regions $(1)$ and $(2)$, which have a rectangular WG cross-section, the WG widths are $a_1$ and $a_2$ respectively (see inset of \cref{WG_drawing}). Since the WG is single-mode and we focus on H-plane bends, these height and widths satisfy $b<\lambda/2$ and $\lambda/2<a_1,a_2<\lambda$, respectively. Inherently, this formalism applies to both symmetric ($a_1=a_2$) and asymmetric ($a_1 \neq a_2$) WG bends. 

The WG is excited from the input port by the fundamental transverse electric (TE) mode $\mathrm{TE}_{10}$, propagating in region $(1)$ towards the junction $(j)$. The operating frequency is $\omega$, and a time harmonic dependence $e^{j\omega t}$ is assumed and omitted from the equations for brevity. As said, a subwavelength polarizable element is placed in the junction as the means to eliminate the reflection loss. In the absence of a scatterer, the junction discontinuity gives rise to scattered fields, required to meet the boundary conditions at the bend. When the scatterer is introduced, induced currents will produce secondary fields, which will contribute to these scattered fields as well.

As in \cite{Widarta1994,Kuwano2003}, we use Cartesian coordinate systems in regions $(1)$ and $(2)$ to conveniently express the fields in rectangular WG eigenmode spectra. These coordinate systems are denoted by $(x_1,y_1,z_1)$ and $(x_2,y_2,z_2)$ and have their origins at points $A$ and $B$ respectively (see \cref{WG_drawing}). In the junction $(j)$, we use a cylindrical coordinate system $(r,\phi,y_j)$ with its origin at the WG corner (point $O$), such that the field there consists of radial modes. We denote the scatterer location in the junction as $(r=r_0,\phi=\phi_0)$.

To avoid limiting the methodology to predefined scatterer geometries, we follow the approach taken in MG analysis and synthesis \cite{PhysRevApplied.8.054037,PhysRevLett.119.067404}, and model the meta-atom as a (virtual) current source. This approach is suitable for subwavelength polarizable elements, where the scattered fields can be well described as emanating from a uniform current \emph{induced} on the scatterer when excited by the incident fields \cite{balanis2012advanced,Osipov2017}. Therefore, we assume that the scatterer can be replaced by a uniform current distribution along the vertical axis ($y_j$), practically forming a 2D problem ($\partial/{{\partial}y}=0$)\footnote{Since the scatterer shape is arbitrary, it may not be uniform along the vertical axis; however, in this case we still assume that since \textcolor{black}{$b$ is substantially small with respect to the wavelength,} 
the spatial variations in current through the scatterer do not give rise to \textcolor{black}{significant (i.e., slowly-evanescent)} modes featuring $\partial/{{\partial}y}\neq0$. \textcolor{black}{This assumption is also closely related to the assumption used in other metagrating-related work \cite{PhysRevLett.119.067404}, \cite{PhysRevApplied.8.054037}, and \cite{Popov2020}, where loaded (generally non-uniform) wires are successfully modeled as uniform current-carrying line sources.} \textcolor{black}{Following this rationale,} we keep treating the problem as 2D, and indeed, as shall be shown in \cref{subsec:passive_scatterer_assignment} (which includes case-studies with vertically non-uniform scatterers), \textcolor{black}{the results obtained with this approximation incorporated into the semianalytical model are in excellent agreement with those recorded in full-wave simulations.}}. 
With this in mind, the entire electric field $\mathbf{E}$ and magnetic field $\mathbf{H}$ along the structure satisfy $\mathbf{E}\times\mathbf{\hat{y}}=\mathbf{H}\cdot\mathbf{\hat{y}}=0$, where $\mathbf{\hat{y}}$ is a unit vector parallel to the narrow WG walls (the walls of length $b$). Consequently, the scattered fields propagating from the junction towards the input and output ports consist of a superposition of $\mathrm{TE}_{n0}$ modes ($n \in \mathbb{N}$), where all modes for $n>1$ are evanescent \cite{balanis2012advanced}.

Since we model the scatterer as a current line placed at an arbitrary point in the junction, then for given WG dimensions and bend angle, the suggested configuration features three DOFs: the scatterer radial distance ($r_0$) and azimuthal offset ($\phi_0$) with respect to the junction coordinate system origin $O$, as well as the current flowing through the line, $I$. This current is clearly related to the meta-atom polarizability, stemming from its yet-to-be-determined geometry, and therefore, from now on it will be referred to as "the induced current"; the field radiated from it will be referred to as "the secondary field". In the following sections, we will derive the detailed synthesis procedure, prescribing how these DOFs can be set to satisfy the design goals. As we lay out the methodology, the implications of the number of available DOFs, as well as the possibilities to implement the meta-atom in practice, will be discussed and illustrated.

\subsection{Field Representation}
\label{subsec:field_representation}
To find the scattering parameters using mode-matching, we first need to have a modal representation of the fields in each of the three regions in \cref{WG_drawing}. We begin by expressing the electric fields in regions $(1)$ and $(2)$ as superpositions of rectangular WG modes, written in their corresponding coordinate systems, $(x_1,y_1,z_1)$ and $(x_2,y_2,z_2)$ respectively (\cref{WG_drawing}). The field in the input region $(1)$ consists of the incident $\mathrm{TE}_{10}$ mode \cite{balanis2012advanced},
\begin{equation}
\label{eq1}
\mathbf{E}_\mathbf{inc} = \mathbf{\hat{y}} E_{\mathrm{in}}\sin{\left(\frac{\pi x_1}{a_1}\right)} e^{j\beta^{(1)}_1 z_1},
\end{equation}
and the reflected $\mathrm{TE}_{n0}$ modes,
\begin{equation}
\label{eq2}
\mathbf{E}_\mathbf{r}=\mathbf{\hat{y}} \sum_{n=1}^{\infty}(A_{n}+A^{\mathrm{(sec)}}_{n})\sin{\left(\frac{n\pi x_1}{a_1}\right)} e^{-j\beta^{(1)}_n z_1},
\end{equation}
where $A_{n}$ are the amplitudes associated with the field reflected from the bend in the absence of the scatterer (hereafter the "bare junction" configuration), and $A^{\mathrm{(sec)}}_{n}$ are the amplitudes corresponding to the secondary field due to the induced current $I$. Similarly, the field in the output region $(2)$ is given by
\begin{equation}
\label{eq3}
\mathbf{E}_\mathbf{t}=\mathbf{\hat{y}} \sum_{n=1}^{\infty}(B_{n}+B^{\mathrm{(sec)}}_{n})\sin{\left(\frac{n\pi x_2}{a_2}\right)} e^{j\beta^{(2)}_n z_2},
\end{equation}
where $B_{n}$ are the amplitudes associated with the field transmitted from the bend in the "bare junction" configuration, and $B^{\mathrm{(sec)}}_{n}$ are the amplitudes corresponding to the secondary field. Using the wavenumber $k=\omega/c$, the propagation constants in \crefrange{eq1}{eq3} are defined as
\begin{equation}
\label{eq4}
\beta^{(q)}_n = 
    \begin{dcases}
      -j\sqrt{\left(\frac{n\pi}{a_{q}}\right)^2-k^2} & k < \frac{n\pi}{a_{q}}\\
      \sqrt{k^2-\left(\frac{n\pi}{a_{q}}\right)^2} & k \geq \frac{n\pi}{a_{q}}
    \end{dcases},
\end{equation}
where $q\in \{1,2\}$, satisfying both the wave equation and the radiation condition.

As mentioned in the previous subsection, we treat the junction region $(j)$ as a section of a radial WG \cite{Widarta1994,Kuwano2003}. In particular, since the scatterer is situated in the junction, the field $\mathbf{E}_\mathbf{j}$ there should satisfy the inhomogeneous Helmholtz equation
\begin{equation}
\label{eq5}
\left(\mathbf{\nabla}^2+k^2\right)\mathbf{E}_\mathbf{j}=\mathbf{\hat{y}} jk\eta \frac{I}{r_0} \delta(r-r_0)\delta(\phi-\phi_0),
\end{equation}
where $\eta=\sqrt{\mu_0/\epsilon_0}$ is the free-space wave impedance, $(r,\phi)$ is the observation point within the junction region $(j)$, and recalling that the current-carrying scatterer is modelled, under our assumptions, as an infinitesimal line source. The solution to this equation consists of a source term as well as a homogeneous term, taking into consideration the input $\mathrm{TE}_{10}$ excitation and multiple scattering in the bend (i.e. guaranteeing satisfaction of the boundary conditions on the radial WG walls). The source term corresponds to the electric field excited by a current line parallel to a conducting wedge\footnote{Since $\partial/{\partial y} = 0$, the solution for a conducting wedge is equivalent to the solution for a radial WG.} \cite{Osipov2017},
\begin{equation}
\label{eq6}
\mathbf{E}^{\mathbf{(s)}}_\mathbf{j}= - \mathbf{\hat{y}} k\eta I \frac{\pi}{\Phi} \sum_{\mu \in M}{J_\mu \left(kr_<\right)H_\mu^{(2)}\left(kr_>\right)\sin{\left(\mu \phi_0\right)}\sin{\left(\mu \phi\right)}},
\end{equation}
where $M=\{m\pi/{\Phi} \mid m\in\mathbb{N}\}$, $r_<=\min{\left(r,r_0\right)}$ and $\ r_>=\max{\left(r,r_0\right)}$. The modal representation of the homogeneous term consists of radial modes, involving only Bessel functions of the first kind \cite{Widarta1994,Kuwano2003},
\begin{equation}
\label{eq7}
\mathbf{E}^{\mathbf{(h)}}_\mathbf{j}=\mathbf{\hat{y}} \sum_{\mu \in M} C_\mu J_\mu \left(kr\right)\sin{\left(\mu \phi\right)},
\end{equation}
where $C_{\mu}$ are the unknown amplitudes of the radial modes composing the homogeneous term. Once they are resolved, the entire field in the junction can be written as $\mathbf{E}_\mathbf{j}=\mathbf{E}^{\mathbf{(h)}}_\mathbf{j}+\mathbf{E}^{\mathbf{(s)}}_\mathbf{j}$.

\subsection{Mode Matching}
\label{subsec:mode_matching}
To suppress the fields reflected from the bend, as desired herein, we wish to enforce constraints on the system scattering parameters. As the latter are directly related to the unknown mode amplitudes $A_1$, $A_1^{\mathrm{(sec)}}$, $B_1$ and $B_1^{\mathrm{(sec)}}$, our next goal is to retrieve these using the analytical model. Specifically, after having written the fields in each of the three regions of the structure in their corresponding modal expansions, as given in \crefrange{eq1}{eq3}, \cref{eq6} and \cref{eq7}, we perform mode-matching on the interfaces between region $(j)$ and regions $(1)$ and $(2)$, namely on lines $\overline{AC}$ and $\overline{BC}$ respectively \cite{Widarta1994,Kuwano2003}. To simplify the solution, we restrict the scatterer location such that it satisfies
\begin{equation}
\label{eq9}
{r_0} \leq \min(h_1,h_2),
\end{equation}
where 
\begin{equation}
\begin{dcases}
    h_1 = AO = \frac{a_2+a_1\cos(\Phi)}{\sin(\Phi)} \\
    h_2 = BO = \frac{a_1+a_2\cos(\Phi)}{\sin(\Phi)},
\end{dcases}\label{eq8}
\end{equation}
such that $\overline{AO}$ and $\overline{BO}$ are perpendicular to $\overline{AC}$ and $\overline{BC}$, respectively, $C$ being the inner corner of the bend (\cref{WG_drawing}). Under this restriction, observation points on the mode-matching lines always satisfy $r \geq r_0$, thus enabling a more convenient formulation of \cref{eq6}. Consequently, we define ${\zeta_\mu=-(k\eta \pi / \Phi) J_\mu(kr_0)\sin(\mu \phi_0)}$ and write the electric field given by \cref{eq6} and \cref{eq7} on the mode-matching lines, in the junction coordinate system $(r,\phi)\in\overline{AC},\overline{BC}$, as
\begin{equation}
\begin{split}
\label{eq10}
\!\!\!\!\!\mathbf{E}^\mathbf{MM}_\mathbf{j} = \mathbf{\hat{y}} \!\sum_{\mu \in M}\!\left(C_\mu J_\mu \left(kr\right)\! +\! I\zeta_\mu H_\mu^{(2)}\left(kr\right)\right)  \sin{\left(\mu \phi\right)}.
\end{split}
\end{equation}

The mode-matching procedure involves equating the fields in region $(1)$ and $(j)$ at the interface $\overline{AC}$, and the fields in region $(2)$ and $(j)$ at the interface $\overline{BC}$. We then use the orthogonality of the rectangular modes to resolve the $n$th mode ($\mathrm{TE}_{n0}$) amplitudes in each rectangular region [see \cref{eq2} and \cref{eq3}], by summing the projections of the radial modes in the junction on the corresponding rectangular mode, given as coupling integrals. Therefore, the mode-matching solution can be written (as shall be shown promptly) in matrix form to be solved via standard matrix inversion. In practice, to form a solvable set of linear equations, the infinite sums in \cref{eq2}, \cref{eq3} and \cref{eq10} should be truncated such that $N$ terms are retained in the expression for $\mathbf{E}_\mathbf{r}$ and $\mathbf{E}_\mathbf{t}$, and $2N$ terms in the summation associated with $\mathbf{E}^\mathbf{MM}_\mathbf{j}$ \cite{Widarta1994}.

The solution for the amplitudes of the $\mathrm{TE}_{n0}$ modes in the input and output ports ($A_n$, $A_n^{\mathrm{(sec)}}$, $B_n$, $B_n^{\mathrm{(sec)}}$), and for the amplitudes of the radial modes of the homogeneous term in the junction ($C_\mu$), is thus given by
\begin{equation}
\begin{cases}
    A_n = -E_{\mathrm{in}} \delta_{1n} + [R_{n\mu,1}^{(1)}][\Lambda_1]^{-1}[A_\mathrm{inc}] \\
    A_n^{\mathrm{(sec)}} = I\left([R_{n\mu,2}^{(1)}] - [R_{n\mu,1}^{(1)}][\Lambda_1]^{-1}[\Lambda_2]\right)[\zeta_\mu] \\
    B_n = [R_{n\mu,1}^{(2)}][\Lambda_1]^{-1}[A_\mathrm{inc}] \\
    B_n^{\mathrm{(sec)}} = I\left([R_{n\mu,2}^{(2)}] - [R_{n\mu,1}^{(2)}][\Lambda_1]^{-1}[\Lambda_2]\right)[\zeta_\mu] \\ 
    C_\mu = [\Lambda_1]^{-1}[A_\mathrm{inc}] - I[\Lambda_1]^{-1}[\Lambda_2][\zeta_\mu],
\end{cases}
\label{eq11}
\end{equation}
where $\delta_{1n}=1$ for $n=1$ and zero otherwise, $[A_\mathrm{inc}] = [E_{\mathrm{in}}, 0 \dots 0]_{1\times 2N}^T$ is the input vector, and the matrices $[\Lambda_p]$ ($p \in \{1,2\}$) are defined as
\begin{equation}
    [\Lambda_p] =\begin{bmatrix}
            \frac{1}{2}(R_{n\mu,p}^{(1)} - Q_{n\mu,p}^{(1)})\\
            R_{n\mu,p}^{(2)} - Q_{n\mu,p}^{(2)}
            \end{bmatrix}_{2N\times 2N} \label{eq12}.\\
\end{equation}

In \cref{eq11} and \cref{eq12}, $R_{n\mu,p}^{(q)}$ and $Q_{n\mu,p}^{(q)}$ are the coupling integrals between the eigenmodes of the $(q)$ and $(j)$ regions ($q\in{1,2}$), reading
\begin{align}
    &R_{n\mu,p}^{(q)} = \frac{2}{a_q} \int_{0}^{a_q} \sin\left(\frac{n\pi x_q}{a_q}\right) \Psi_{p\mu} (kr_q) \sin(\mu \phi_q)dx_q \label{eq13} \\
    &\begin{aligned}
      Q_{n\mu,p}^{(q)} = \frac{2jk}{a_q\beta_n^{(q)}} \int_{0}^{a_q}&\sin\left(\frac{n\pi x_q}{a_q}\right) \left[\frac{h_q}{r_q}\Psi'_{p\mu} (kr_q) \sin(\mu \phi_q)\right. \\
      &+(-1)^q \left.\frac{\mu x_q}{kr_q^2} \Psi_{p\mu} (kr_q) \cos(\mu \phi_q)\right]dx_q       
    \end{aligned} \label{eq14}
\end{align}
where $r_q = \sqrt{x_q^2+h_q^2}$, $\phi_1 = \arctan\left({x_1/h_1}\right)$, $\phi_2 = \Phi-\arctan\left({x_2/h_2}\right)$ are the polar coordinates about $O$ along the mode-matching integration lines, and
\begin{equation}
\Psi_{p\mu}\left(kr_q\right) = 
    \begin{dcases}
      J_\mu \left(kr_q\right) & p=1\\
      H_\mu^{(2)}\left(kr_q\right) & p=2
    \end{dcases}, \label{eq15}\\
\end{equation}
associated with the two different contributions in \cref{eq10}.

Finally, the fractions of power reflected to the input port and transmitted to the output port are given by the squared magnitudes of the bend configuration S-parameters, $S_{11}$ and $S_{21}$, respectively:
\begin{subequations}
\begin{align}
    &{\left|S_{11}\right|}^2 = {\left|\frac{A_1+A_1^{\mathrm{(sec)}}}{E_{\mathrm{in}}}\right|}^2, \label{eq16a}\\
    &{\left|S_{21}\right|}^2 = \gamma^2{\left|\frac{B_1+B_1^{\mathrm{(sec)}}}{E_{\mathrm{in}}}\right|}^2, \label{eq16b}
\end{align}\label{eq16}
\end{subequations}

where
\begin{equation}
    \gamma = \left(\frac{{\left(ka_2\right)}^2 - \pi^2}{{\left(ka_1\right)}^2 - \pi^2}\right)^{\frac{1}{4}} \label{eqGamma}
\end{equation}

takes into account the different wave impedances observed for guided modes propagating in rectangular WGs of different cross sections.

\subsection{Formulation of Constraints}
\label{subsec:constraints}
As mentioned in Subsection \ref{subsec:system_configuration}, the scattering parameters in \cref{eq16} depend on the configuration's three DOFs: the scatterer location $(r_0,\phi_0)$ and the induced current $I$. To enable fulfillment of the design goals, namely, eliminating reflections while using a realistic (i.e. passive) configuration, these DOFs should be set to satisfy two constraints. First, the reflection coefficient must vanish. Namely, the field reflected from the bare junction and the $\mathrm{TE}_{10}$ mode of the secondary field in the input port should destructively interfere \cite{PhysRevLett.119.067404}. In terms of the reflection coefficient, this translates to
\begin{equation}
    {\left|S_{11}\right|}^2 = 0. \label{eq18a}
\end{equation}
Second, we should guarantee that the induced current required to satisfy (17) does not violate the power balance in the system, so that the desired secondary fields can be eventually produced using a passive scatterer. Therefore, and assuming the scatterer is lossless, unitary transmission (power conservation) is additionally demanded, formulated as
\begin{equation}
    {\left|S_{21}\right|}^2 = 1. \label{eq18b}
\end{equation}

These constraints form two nonlinear equations, stipulating relations the meta-atom parameters should satisfy to enable implementation of a seamless bend transition by passive means. Since the chosen device configuration features three DOFs, with redundancy in view of the number of constraints, we expect multiple valid solutions (for given WG dimensions and bend angle). In particular, as shall be demonstrated in Section \ref{sec:resultsanddiscussion}, \cref{eq18a} and \cref{eq18b} will be used to identify locations $(r_0,\phi_0)$ in the junction where a passive solution can be obtained; for each of these, a certain induced current [given by substitution back in \cref{eq11}] would guarantee optimal performance. In the following section we will investigate the properties of these suitable locations, and discuss practical realizations of the scatterer that would facilitate the required induced current.

\section{Results and Discussion}
\label{sec:resultsanddiscussion}

\subsection{Scatterer Location}
\label{subsec:scatterer_location}

As mentioned in the previous section, for a given bend configuration, characterized by certain $a_1$, $a_2$ and $\Phi$, we begin the design process by finding all possible locations $(r_0,\phi_0)$ of the scatterer that satisfy the constraints \cref{eq18a} and \cref{eq18b}. These points are referred to as perfect transmission locations, or PTLs.

To find out if a point $(r_0,\phi_0)$ [restricted by \eqref{eq9}] is a PTL, we first use it in \cref{eq18a}, the constraint dictating zero reflection. Since \cref{eq18a} is equivalent to the destructive interference condition ${A_1+A_1^{\mathrm{(sec)}}=0}$ via \cref{eq16a}, we can use \cref{eq11} to obtain a value for the current $I_\mathrm{NR}$, required to be induced on a meta-atom situated at this location to suppress the WG back-reflections,
\begin{equation}
    I_\mathrm{NR} = - \frac{A_1}{\left([R_{1\mu,2}^{(1)}] - [R_{1\mu,1}^{(1)}][\Lambda_1]^{-1}[\Lambda_2]\right)[\zeta_\mu]},\label{eq19}
\end{equation}

Since at this point we have put no limitations on the way the current $I_\mathrm{NR}$ is excited, suppressing reflections as per \cref{eq19} may involve power absorption or active gain, both undesired from a practical perspective. \textcolor{black}{To make sure that this does not happen, the integral of the real part of Poynting vector around the abstract current source should vanish, which will happen if the current does not disturb the power balance in the system, namely ${\left|S_{11}\right|}^2+{\left|S_{21}\right|}^2=1$, or ${\left|S_{21}\right|}^2=1$ when the return loss is zero. This would ensure that the prescribed scattered field distribution can be reproduced with a purely-reactive element, meeting our goal of a completely passive solution.} Therefore, subsequently, we use $(r_0,\phi_0)$ and $I_\mathrm{NR}$ to calculate ${\left|S_{21}\right|}^2$, and if \eqref{eq18b} is satisfied, it means that the chosen point is indeed a PTL. In other words, if the calculated current $I_\mathrm{NR}$ is induced on a scatterer (or otherwise excited) at the point $(r_0,\phi_0)$, the bend system would exhibit zero reflection and unitary transmission, as desired.

\begin{figure}[!t]
\centerline{\includegraphics[width=\columnwidth]{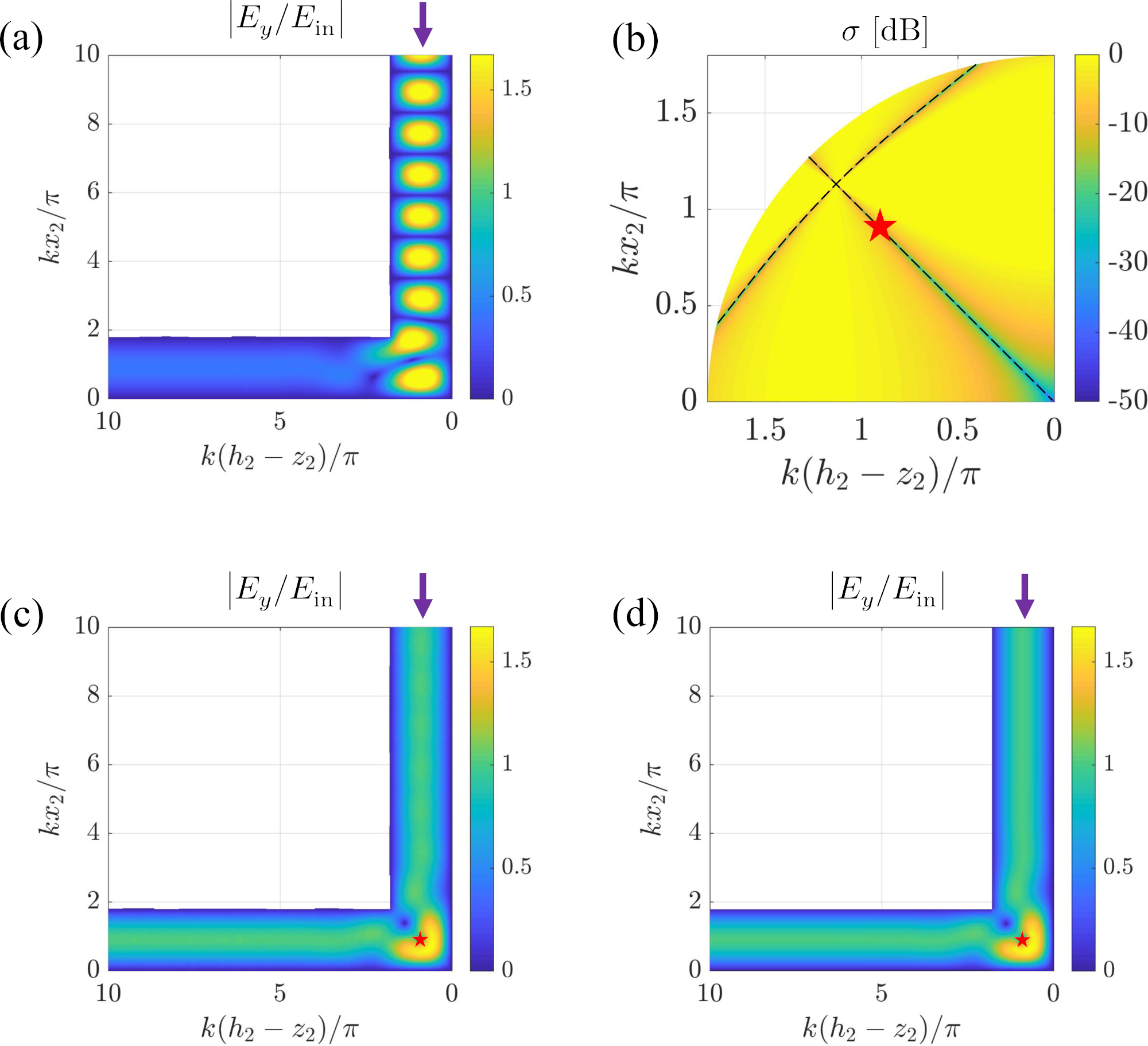}}
\caption{Mitigating reflection loss in a WG bend ($\Phi=90^{\circ}$, $a_1=a_2=0.9\lambda$) using a current-carrying line source (\cref{subsec:scatterer_location}). (a) Electric field magnitude $\left|E_y/E_\mathrm{in}\right|$ as obtained from full-wave simulations for the bare junction. (b) $\sigma$-map of the junction in dB, calculated via \cref{eq20}; dashed lines indicate regions of low deviation from constraints, red star marking the chosen PTL for placing the current line. (c) Full-wave simulation result and (d) semianalytical prediction (\cref{sec:theory}, $N=5$) of the fields with the current line embedded as prescribed in (b) with \cref{eq19}. In all field plots, the purple arrow indicates the input port.}
\label{fig2}
\end{figure}

To demonstrate this PTL detection procedure and verify our semianalytical model, we consider a symmetric ($a_1=a_2=0.9\lambda$) right-angle ($\Phi=90^\circ$) bend, operating at $f=10\mathrm{GHz}$ ($\lambda \approx 3 \mathrm{cm}$), with the WG height being $b=0.25\lambda$. This configuration was chosen in particular, since the analytical model predicted very high reflection loss ($>80\%$) in the bare junction \cite{Widarta1994}. This prediction is corroborated by the full-wave simulation results presented in \cref{fig2}(a) (ANSYS HFSS), where the interference between the incident $\mathrm{TE}_{10}$ mode and the field back-scattered from the junction is clearly observed at the input port.

To find the PTLs for this scenario, we calculate $I_\mathrm{NR}$ via \cref{eq19} for all points $(r_0,\phi_0)$ consistent with \cref{eq9}, and use \cref{eq16b} to compute the deviation from power conservation $\sigma(r_0,\phi_0)$, formally defined as
\begin{equation}
    \sigma(r_0,\phi_0) = \left|1-{\left|S_{21}\right|}^2\right|\bigg|_{{\left|S_{11}\right|}^2=0} .\label{eq20}
\end{equation}
Plotting this quantity as a function of potential scatterer location in the junction forms a convenient visual representation ($\sigma$-map) of the expected deviation from the design goals \cref{eq18a} and \cref{eq18b}, from which the PTLs can be inferred. For the case considered, where \cref{eq9} yields a quarter of a circle as the area to search in for PTLs, this $\sigma$-map is shown in \textcolor{black}{$\mathrm{dB}$} unit scale in \cref{fig2}(b), where we used $N=5$ as the number of modes when calculating \cref{eq11}. The yellow (bright) areas in the map represent regions of large deviation, whereas the dashed lines mark (dark) regions in which the deviation is low ("solution branches"), indicating the PTLs.

As per the PTL identification procedure laid out above, the points on these solution branches may potentially be used to place a scatterer which, if properly excited, would facilitate perfect transmission across the bend. For instance, we may choose here the point $r_0=0.9\lambda/\sqrt{2}, \phi_0=45^{\circ}$ [red star in \cref{fig2}(b)]; according to the semianalytical methodology laid out in \ref{sec:theory}, if we somehow manage to excite the current $I_\mathrm{NR} = 1.94e^{2.69j}E_{\mathrm{in}}b/\eta$, given by \cref{eq19} at this PTL, all the incident power should couple to the output port, without unwanted reflections.

To validate this prediction, we introduced into the bare junction defined in ANSYS HFSS [\cref{fig2}(a)] a line source at the chosen $(r_0,\phi_0)$ carrying the specified current $I_\mathrm{NR}$, and resimulated the modified configuration. The resultant field snapshot is presented in \cref{fig2}(c). Indeed, it can be clearly observed that placing the suitable current in the prescribed location in the junction entirely eliminates the reflections due to the bend, leading to full transmission. \cref{fig2}(d) shows the semianalytical prediction of the electric field for the same configuration, calculated in Matlab using \crefrange{eq1}{eq15} ($N=5$). The excellent agreement between the two field distributions verifies the accuracy of the analytical model, and demonstrates the efficacy of the proposed solution\textcolor{black}{.}

\subsection{Solution Branches - Physical Interpretation}
\label{subsec:branches}
\begin{figure}[!t]
\centerline{\includegraphics[width=\columnwidth]{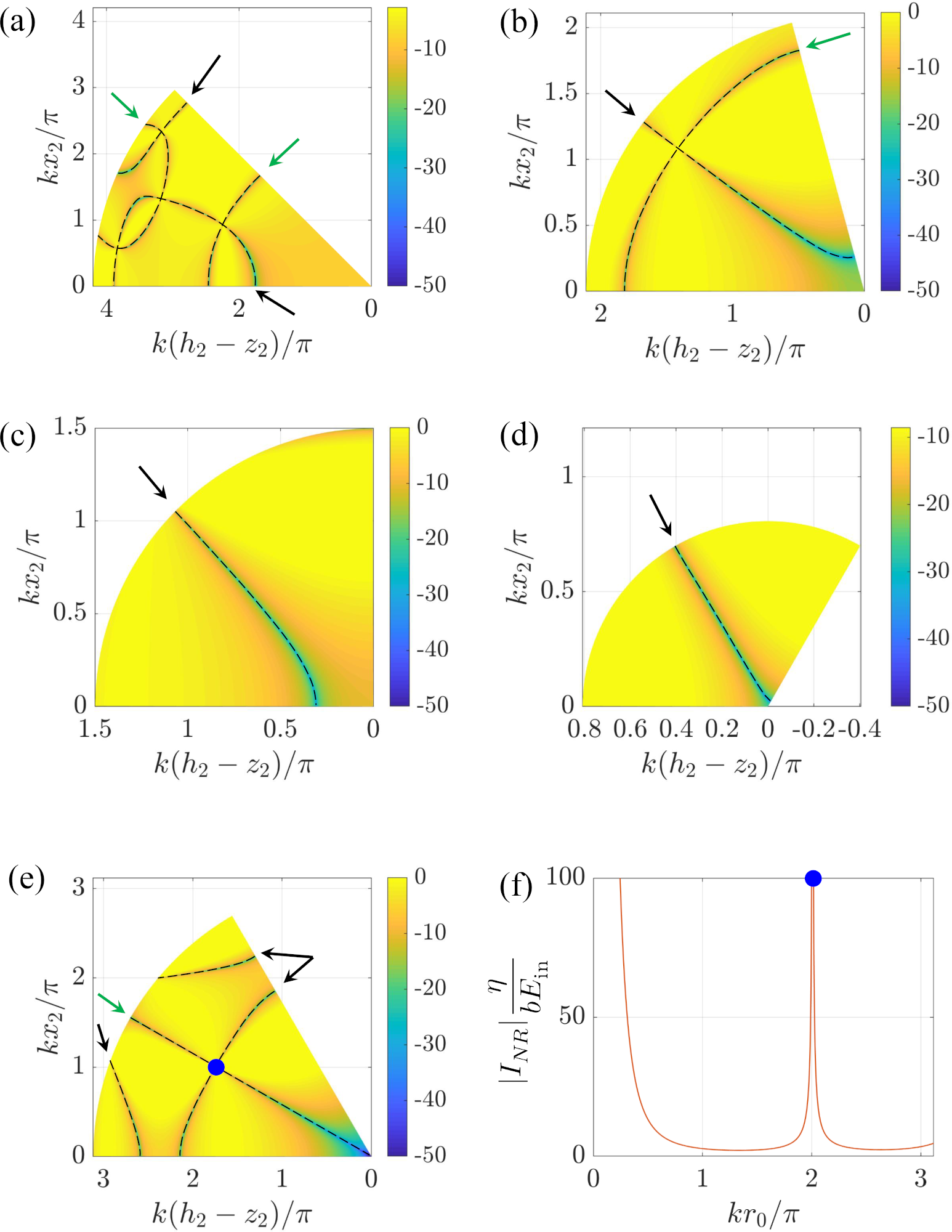}}
\caption{$\sigma$-maps (in $\mathrm{dB}$) for various bend configurations, calculated via \cref{eq20} with $N=5$ (dashed lines indicate solution branches). (a) $\Phi = 45^{\circ}, a_1 = 0.9\lambda, a_2 = 0.85\lambda$ (b) $\Phi = 75^{\circ}, a_1 = 0.85\lambda, a_2 = 0.8\lambda$ (c) $\Phi = 90^{\circ}, a_1 = 0.75\lambda, a_2 = 0.95\lambda$ (d) $\Phi = 120^{\circ}, a_1 = 0.8\lambda, a_2 = 0.75\lambda$ (e) $\Phi = 60^{\circ}, a_1 = a_2 = 0.9\lambda$. (f) The (normalized) required induced current needed for each point in the axis of symmetry of configuration (e), where the blue circle corresponds to the blue circle in (e). The black and green arrows mark the symmetric and anti-symmetric branches, respectively.}
\label{fig3}
\end{figure}

Before we discuss options to synthesize practical reflectionless WG bends (i.e. without the use of impressed current sources as in the example presented in Section \ref{subsec:scatterer_location}) based on the outlined approach, we wish to examine typical features of the PTLs, leading to physical insights which may assist in choosing appropriate PTLs. To this end, we plot in \cref{fig3}(a)-(e) $\sigma$-maps associated with WG junctions of various bending angles (from $45^{\circ}$ to $120^{\circ}$) and WG widths (including cases of $a_1 \neq a_2$ and a case of $a_1=a_2$). As can be seen, in all the considered configurations, the PTLs are located on continuous curves (solution branches), which in some cases intersect [this feature was also previously observed in \cref{fig2}(b)].

We begin by investigating PTLs from an analytical perspective. In particular, we first introduce a sufficient condition under which a point $(r_0,\phi_0)$ is a PTL, stating that the \emph{secondary} $\mathrm{TE}_{10}$ fields radiated from the current excited there satisfy
\begin{equation}
    A_1^{\mathrm{(sec)}} = \pm je^{j\xi} \gamma B_1^{\mathrm{(sec)}},\label{eq21}
\end{equation}
where $\xi$ is defined as the phase difference between the \emph{bare junction} reflected and transmitted propagating modes, namely $\xi =\arg{\left(B_1^* A_1\right)}$. To show that this is indeed a sufficient condition for a PTL, we must prove that a scatterer location $(r_0,\phi_0)$ obtained from it\footnote{Note that \cref{eq21} is independent of the current, since the secondary fields depend on $I$ linearly [see \cref{eq11}], and we clearly assume that $I\neq0$. Therefore, the solution to \cref{eq21} is a collection of scatterer locations $(r_0,\phi_0)$ in the junction.} satisfies both \cref{eq18a} and \cref{eq18b} \emph{using the same induced current}. This is equivalent to showing that the current cancelling the back-reflections, $I_{\mathrm{NR}}$, obtained at this location, solves the power-balance constraint \cref{eq18b} [this equivalence stems from the fact that $I_{\mathrm{NR}}$ is the \emph{only} current solving \cref{eq18a}]. This can be done easily by plugging \cref{eq21} into \cref{eq16b}, and using \cref{eq11} and \cref{eq19} to obtain the equation ${|A_1|^2 + |\gamma B_1|^2 = |E_{\mathrm{in}}|^2}$, which is always true, since this is the equation of power conservation in the (passive) bare junction.

While the relation \cref{eq21} might look somewhat obscure, it can actually be rationalized using the physical reasoning underlying its rigorous proof discussed above. Indeed, let us assume that we managed to excite the reflection-cancelling current, $I_{\mathrm{NR}}$, in some location $(r_0,\phi_0)$ solving \cref{eq21}, such that destructive interference takes place in the input port (${A_1 = -A_1^{\mathrm{(sec)}}}$). Furthermore, since $\gamma$ considers the different wave impedances of the input and output ports, \cref{eq21} implies that the secondary $\mathrm{TE}_{10}$ fields emanating from the scatterer and propagating towards the output port ($B_1^{\mathrm{(sec)}}$) and the input port ($A_1^{\mathrm{(sec)}}$) carry exactly the same power. Hence, when \cref{eq21} holds and $A_1 = -A_1^{\mathrm{(sec)}}$, the power carried by $B_1^{\mathrm{(sec)}}$ equals the power carried by $A_1$. We already know, as stated in the proof above, that ${|A_1|^2 + |\gamma B_1|^2 = |E_{\mathrm{in}}|^2}$, and \textcolor{black}{by substituting \cref{eq16b} into \cref{eq18b}, the power-balance constraint can be written as} ${\gamma^2|B_1 + B_1^{\mathrm{(sec)}}|^2 = |E_{\mathrm{in}}|^2}$.
Since $A_1$ and $B_1^{\mathrm{(sec)}}$ carry exactly the same power, \textcolor{black}{then to guarantee that the latter equation is satisfied and power balance is attained,} $B_1^{\mathrm{(sec)}}$ and $B_1$ must not interfere 
(otherwise, the non-zero interference term will violate the power balance). To achieve that, $B_1^{\mathrm{(sec)}}$ and $B_1$ must have a phase difference of $\pm \pi/2$ (quarter-wavelength) between them. Assuming a phase difference of $\xi$ between the fields scattered from the bare junction\footnote{When discussing $\xi$, we assume that $B_1$ and $A_1$ are not zero, i.e. cases in which $\xi$ is basically undefined. Furthermore, note that when $B_1=0$, \cref{eq16b} will not have any interference term, and therefore this case is irrelevant here. When $A_1=0$, the junction transmits all power to begin with, so it does not concern us \textcolor{black}{either}.}, $A_1$ and $B_1$, this desired quarter-wavelength phase difference will be obtained when the secondary fields have between them a phase difference of $\xi$ appended by an additional phase difference of $\pm \pi/2$. As can be observed, this is the \emph{exact} relation stipulated in \cref{eq21}, meaning that this phase difference is inherently achieved when we choose a point solving \cref{eq21} to begin with. Overall, we can see that once a scatterer is placed at a location obtained via \cref{eq21}, both zero reflection and power balance will be acquired using the same induced current, and therefore \cref{eq21} is a \emph{sufficient} condition for some point to be a valid PTL.

In fact, all the PTLs we encountered so far, including all the case studies presented in Section \ref{sec:resultsanddiscussion} herein, satisfy \cref{eq21}. Moreover, we observed that in all these cases, the set of solutions can be associated with continuous solution branches, with each branch hosting \emph{exclusively} PTLs which satisfy \cref{eq21} with either the plus sign or the minus sign (but not both) on the right-hand-side. Correspondingly, we denote the solutions lying on the branches formed by \cref{eq21} with a plus sign as \emph{symmetric}, while the ones corresponding to \cref{eq21} with a minus sign are referred to as \emph{anti-symmetric} branches, marked, respectively, by black and green arrows in \cref{fig3}(a)-(e).

In symmetric WG bends\footnote{Note the distinction between symmetric WGs and symmetric solution branches, where the former correspond to the geometrical dimensions of the WG and the latter stand for PTL identification.}, i.e. $a_1=a_2$, it can be shown that $|\xi|=\pi/2$ [reducing \eqref{eq21} to $A_1^{\mathrm{(sec)}} = \pm B_1^{\mathrm{(sec)}}$] and that if the current line is placed on the symmetry axis of the bend ($\phi_0=\Phi/2$), the amplitudes of the secondary fields are equal at both ports ($A_1^{\mathrm{(sec)}} = B_1^{\mathrm{(sec)}}$). Namely, the axis of symmetry\footnote{We refer here only to points on the axis of symmetry which are restricted by \cref{eq9}, as considered herein.} is always a branch of the solution when $a_1=a_2$ [as indeed seen in Figs. 2(b) and 3(e)]. This phenomenon can be explained intuitively by the observation that when placed at the structure's geometrical axis of symmetry, the scatterer radiates identical fields to both ports. Therefore, if the bare junction reflection is exactly canceled by the scatterer radiation towards the input port, its radiation towards the output port would fully compensate for the eliminated back-reflected power, leading to full transmission. Therefore, this scatterer location is a PTL.

Another observation concerning the solution branches that can be deduced from \cref{eq21} is related to the nature of the solutions at intersections between symmetric and anti-symmetric branches: the intersection points lie on both branches, and must therefore satisfy \cref{eq21} with \emph{both} plus and minus signs on the right-hand-side. Quite clearly, this can only occur if the secondary fields at those points vanish, i.e. $A_1^{\mathrm{(sec)}} = B_1^{\mathrm{(sec)}} = 0$. Since we are only interested in configurations which exhibit undesired reflection to begin with, i.e. $A_1\neq0$, no finite current excited at those (exact) intersection points will be able to satisfy the design requirement specified in \cref{eq18a}, $A_1^{\mathrm{(sec)}}=-A_1$. Therefore, these intersection points cannot be considered valid PTLs\footnote{Moreover, at these intersection points the value of $\sigma$ is undefined, since its definition in \eqref{eq20} inherently assumes ${\left|S_{11}\right|}^2=0$, which does not occur there when $A_1$ is nonzero.}. Physically, a current source placed in these points in the junction would not be able to generate any $\mathrm{TE}_{10}$ power radiated towards port (1), due to special multiple reflection phenomena occurring there, leading to complete destructive interference in the secondary field pattern. \textcolor{black}{From another perspective, these points are actually “blind spots”: from reciprocity, the bare-junction field there is zero. In other words, the incident TE10 mode would not be able to efficiently excite currents in a polarizable particle placed at these points, and a scatterer there will be “invisible” to the system. In particular, it will not be possible to induce on the scatterer a significant (i.e., sufficient) amount of current to produce the desired secondary fields and cancel the spurious back reflection, as required.}

Naturally, similar effects are experienced by current elements placed in the vicinity of these intersection points. Although this time the destructive interference in the secondary field pattern at the input port is imperfect, it is very significant nonetheless, such that extremely high currents are required to generate substantial field magnitudes, needed to eliminate the bare-junction reflection\footnote{From a mathematical perspective, since \cref{eq21} is independent of the current, the quantity approaching zero near the intersection point (the point where $A_1^{\mathrm{(sec)}} = 0$) is actually $A_1^{\mathrm{(sec)}}/I$; therefore, to compensate for this extremely small value of this quantity, $I$ must be very high to satisfy $A_1^{\mathrm{(sec)}} = -A_1$.}. Therefore, as was also observed and discussed in detail in \cite{PhysRevApplied.8.054037} for beam-manipulating MGs, the points in proximity to these branch intersections are not suitable PTLs: the extreme currents required to be excited would lead to significant power dissipation in any realistic scatterer (which inevitably includes finite, even if small, loss). Thus, in practice, PTLs for placing passive scatterers when realizing the bend transmission solution (see Section \ref{subsec:passive_scatterer_assignment}) should be chosen away from those intersection points. This observation is demonstrated in \cref{fig3}(f), showing the (normalized) required induced current $I_\mathrm{NR}$ as a function of $r_0$ on the axis of symmetry for the configuration of \cref{fig3}(e) ($\phi_0=\Phi/2=30^{\circ}$); as clearly observed, the current diverges as we approach the intersection point of the two branches [blue dot in Figs. 3(e) and 3(f)].

Taking another look at \cref{fig3}(f), one discovers that the current diverges also as $r_0 \rightarrow 0$. Actually, this can already be inferred from \cref{eq19}, where the constant $\zeta_\mu$ in the denominator [defined after \cref{eq9}] can be seen to vanish at this point. The reason for this divergence is quite intuitive, and can be explained using image theory \cite{balanis2012advanced}: when an electric current element is placed in close proximity to a highly-conducting wall, tangential current components would not be able to radiate effectively, due to destructive interference with the field produced by their image. Therefore, once again, the current required to be induced on scatterers positioned in these regions in order to suppress reflections from the bare junction would be enormous, leading, as discussed in the previous paragraph, to substantial loss when practical implementation is considered. As a result, one should also avoid choosing PTLs which are close to the WG walls for obtaining efficient transmission across the bend.

\subsection{Passive Scatterer Assignment}
\label{subsec:passive_scatterer_assignment}

In Section \ref{subsec:scatterer_location}, we have demonstrated how the analytical model developed in Section \ref{sec:theory} can be used to retrieve suitable locations and current amplitudes such that, if an element carrying the prescribed current is placed at the corresponding PTL, reflection loss due to the WG bend will be eliminated, and unitary transmission will be achieved. However, we have yet to indicate how to realize this current excitation in a practical WG system. This challenge was circumvented in Sections \ref{subsec:scatterer_location} by using an impressed current source to verify the validity and accuracy of the proposed semianalytical methodology. Nonetheless, as denoted in Sections \ref{sec:introduction} and \ref{sec:theory}, our ultimate goal is to present a synthesis procedure which would lead to a completely passive configuration to solve the scattering problem at the junction. To this end, similarly to analogous schemes derived for beam-manipulating MGs \cite{PhysRevLett.119.067404,PhysRevApplied.8.054037}, we wish to replace the impressed current source with a suitable passive polarizable particle. The geometry of this scatterer should be judiciously chosen and fine-tuned such that, when excited by the incident $\mathrm{TE}_{10}$ field incoming from the input port, the current induced on it would generate the secondary fields dictated by \cref{eq18a} and \cref{eq18b}, thus guaranteeing perfect power transmission across the bend.

\begin{figure}[!t]
\centerline{\includegraphics[width=\columnwidth]{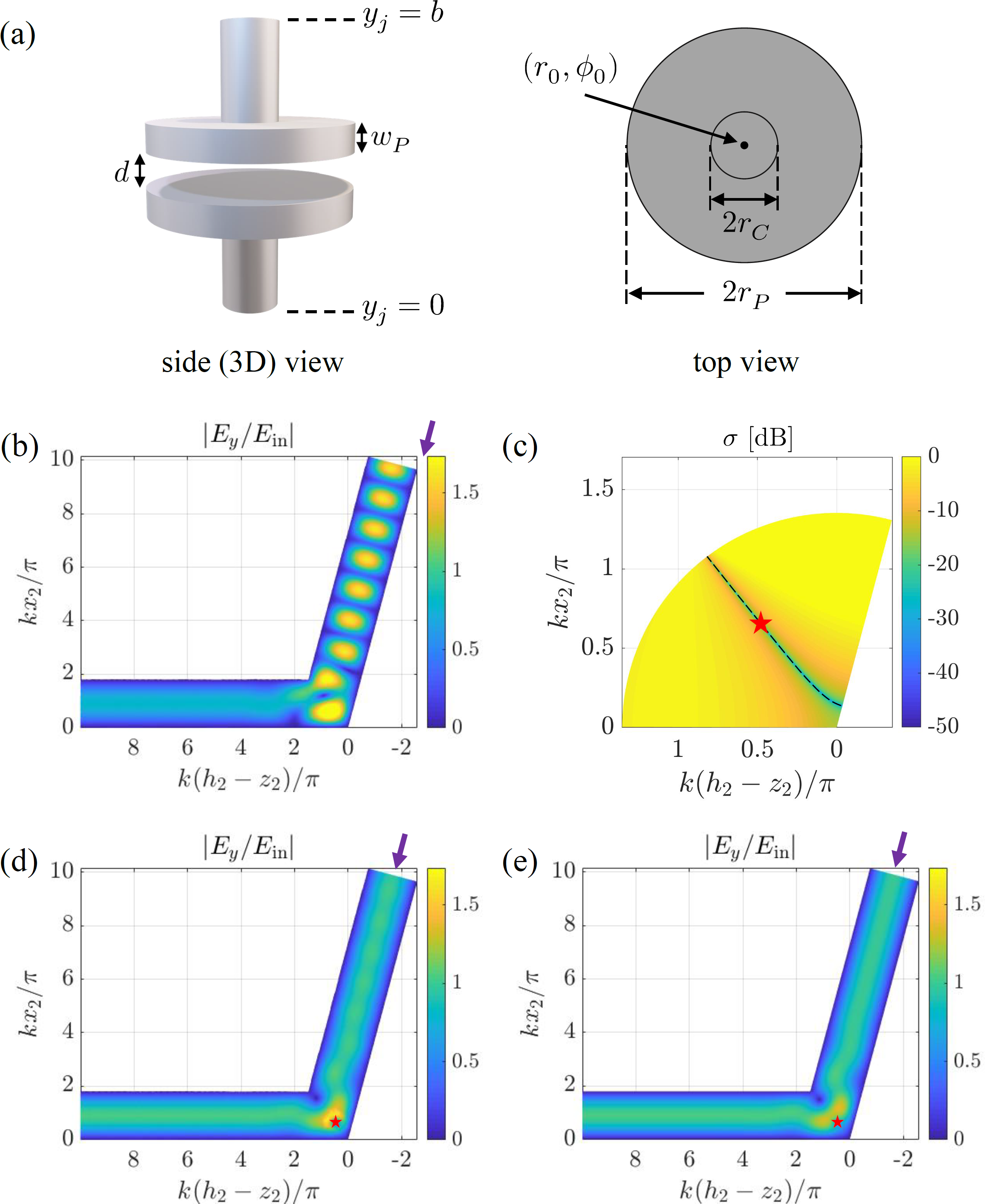}}
\caption{Mitigating reflection loss in an asymmetric WG bend ($\Phi=105^{\circ}$, $a_1=0.95\lambda$, $a_2=0.9\lambda$) using a capacitively loaded wire (\cref{subsec:passive_scatterer_assignment}). (a) Scatterer geometry. 
(b) Electric field magnitude $\left|E_y/E_\mathrm{in}\right|$ in the bare junction, obtained via full-wave simulation. (c) $\sigma$-map of the junction in dB, calculated via \cref{eq20}; dashed lines indicate regions of low deviation from constraints, red star marking the chosen PTL for placing the scatterer. (d) Full-wave simulation result and (e) semianalytical prediction (\cref{sec:theory}, $N=5$) of the fields with the scatterer embedded as prescribed in (c) with $r_P = 0.2\lambda$ and $d = 0.4b$. In all field plots, the purple arrow indicates the input port.}
\label{fig4}
\end{figure}

In this subsection, we demonstrate the feasibility of such a passive solution by using a capacitively-loaded wire as the required electrically polarizable scatterer, in the spirit of our previous work \cite{PhysRevApplied.8.054037,Rabinovich_2,Rabinovich_3,Rabinovich2018}. The meta-atom and its geometrical parameters are shown in \cref{fig4}(a), featuring a cylindrical symmetry, more suitable for the WG bend problem. The wire will be placed at a PTL derived by the semianalytical methodology, while geometrical dimensions of the capacitor will be adjusted using a parametric sweep in a full-wave solver to reach the design point in which the secondary fields emanating from the scatterer yield maximal transmission from input to output. In the following case studies, the scatterer is made of copper, and the wire radius and plate thickness are both fixed at $r_C = w_P = \lambda/50$, while the adjusted parameters are the plate radius $r_P$ and the distance between the plates $d$.

We consider two different WG bend configurations, with a field excitation at $f=10\mathrm{GHz}$ ($\lambda \approx 3\mathrm{cm}$) and WG height of $b=0.25\lambda$, as in Subsection \ref{subsec:scatterer_location}. We first examine an asymmetric $105^{\circ}$ bend, with its input and output WG widths (\cref{WG_drawing}) given by $a_1=0.95\lambda$ and $a_2=0.9\lambda$, respectively. For this configuration, the bare junction suffers more than $40\%$ reflection loss [\cref{fig4}(b)]. To mitigate this undesired scattering, we follow the scheme laid out in Section \ref{sec:theory} and demonstrated in Subsection \ref{subsec:scatterer_location}, and generate the $\sigma$-map presented in \cref{fig4}(c), indicating possible locations for scatterers (dashed black line). Considering the discussion in Subsection \ref{subsec:branches}, we choose to place the scatterer at the PTL marked by a red star on the map, corresponding to the point $(r_0= 0.398\lambda$,$\phi_0=51.13^{\circ})$ in the junction. The WG bend with the capacitively-loaded wire placed at the chosen PTL was defined and simulated in ANSYS HFSS. Sweeping the capacitor plate radius $r_p$ and gap $d$ revealed a number of combinations yielding over $99\%$ transmission; for the purpose of this demonstration, we choose one of these, setting $r_P = 0.2\lambda$ and $d = 0.4b$. A field snapshot for this chosen fully passive configuration as recorded in full-wave simulations is shown in \cref{fig4}(d), indicating excellent agreement with semianalytical predictions obtained by using \cref{eq19} in \cref{eq11} [\cref{fig4}(e)], validating the synthesis procedure.

\begin{figure}[!t]
\centerline{\includegraphics[width=\columnwidth]{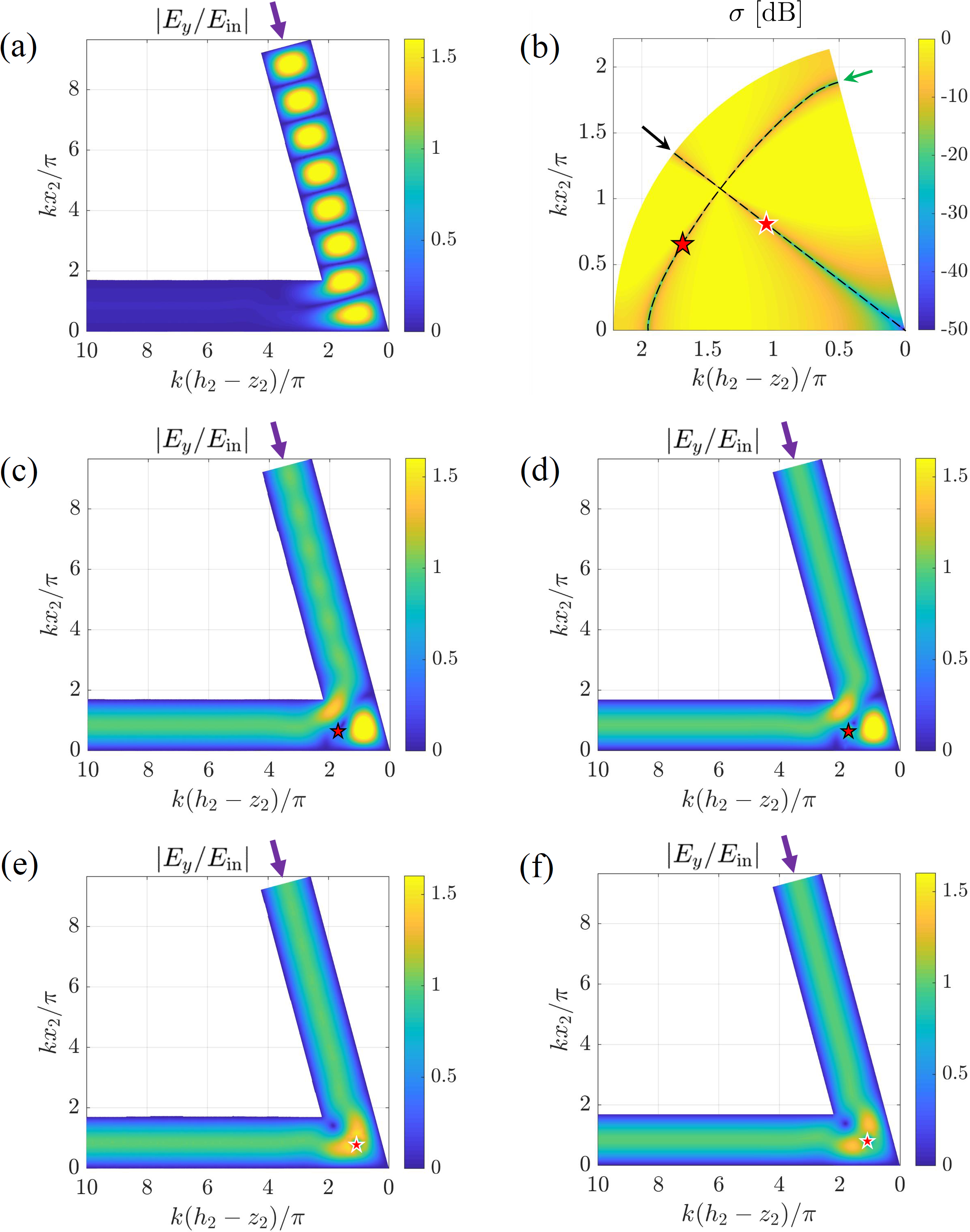}}
\caption{Mitigating reflection loss in a symmetric WG bend ($\Phi=75^{\circ}$, $a_1=a_2=0.85\lambda$) using a capacitively loaded wire as depicted in \cref{fig4}(a) (\cref{subsec:passive_scatterer_assignment}). (a) Electric field magnitude $\left|E_y/E_\mathrm{in}\right|$ in the bare junction, obtained via full-wave simulation. (b) The $\sigma$-map with the chosen PTLs marked with red stars, distinguished by black and white outlines for the anti-symmetric and symmetric solution branches, respectively. (c) Full-wave simulation results and (d) semianalytical prediction (\cref{sec:theory}, $N=5$) of the fields, for the case of the PTL in the anti-symmetric branch (indicated by a green arrow). (e) Full-wave simulation results and (f) semianalytical prediction (\cref{sec:theory}, $N=5$) of the fields, for the case of the PTL in the symmetric branch (indicated by a black arrow). In all field plots, the purple arrow indicates the input port.}
\label{fig5}
\end{figure}

To demonstrate the broad applicability of this general scheme for a wide variety of WG configurations and bend angles, we choose as another example a symmetric $75^{\circ}$ bend, where $a_1=a_2=0.85\lambda$, considered in \cref{fig5}. For this configuration, the situation is even more severe to begin with, as the bare junction scatters back more than $99\%$ of the incident power [\cref{fig5}(a)]. Once again, we follow the methodology prescribed in Section \ref{sec:theory} and Subsection \ref{subsec:scatterer_location}, yielding the $\sigma$-map presented in \cref{fig5}(b). Recalling the discussion in Section \ref{subsec:branches}, we can identify symmetric and anti-symmetric solution branches here, denoted by the black and green arrows, respectively. Even if somewhat unintuitive at first, the formalism implies that points on either branch can be used to obtain perfect transmission. To highlight this observation, we present two different solutions for cancelling the bend reflections in this case, using PTLs from both branches. First, from the anti-symmetric branch we choose the point $(r_0= 0.907\lambda, \phi_0=53.81^{\circ})$ [black-outlined red star in \cref{fig5}(b)]. To determine the load capacitor dimensions [\cref{fig4}(a)], a brief sweep in ANSYS HFSS was conducted as mentioned in the previous paragraph, indicating that a radius of $r_P = 0.05\lambda$ and a gap of $d = 0.25b$ would transmit more than $99\%$ of the power across the bend. Simulated fields depicted in \cref{fig5}(c) confirm this result, in agreement with analytical predictions [\cref{fig5}(d)]. Next, from the symmetric branch we choose $(r_0= 2\lambda/3, \phi_0=37.5^{\circ})$ [white-outlined red star in \cref{fig5}(b)]. This time, the scatterer dimensions would, in general, be different than the ones found at the previous PTL, and the parametric sweep points out $r_P = 0.175\lambda$ and $d = 0.2b$ as a good choice. Simulating this configuration in ANSYS HFSS leads again to over $99\%$ transmission, where the field plot as generated by the full-wave solver with the bend+loaded wire configuration [\cref{fig5}(e)] and the one resulting from the analytical model [\cref{fig5}(f)] match very well.

These examples demonstrate the great design flexibility our approach offers to engineers, allowing selection between multiple possible solutions for a particular bend configuration. As shown, even an extreme case of over $99\%$ reflection in a WG junction can be efficiently mitigated by a single polarizable element using the presented MG-inspired technique.

\subsection{Scatterer Geometry Retrieval}
\label{subsec:cylinder_radius_model}
While the methodology outlined in the previous sections allows efficient semianalytical evaluation of the valid locations for a scatterer enabling perfect transmission across the bend, finding the suitable scatterer geometry still involves parametric sweeps in a full-wave solver, which could become quite time-consuming. Therefore, similarly to \cite{PhysRevApplied.8.054037,Rabinovich_2,Rabinovich2018}, it would be beneficial to devise an extension to the analytical model that would be able to cope with this last synthesis step as well.

\begin{figure}[!t]
\centerline{\includegraphics[width=\columnwidth]{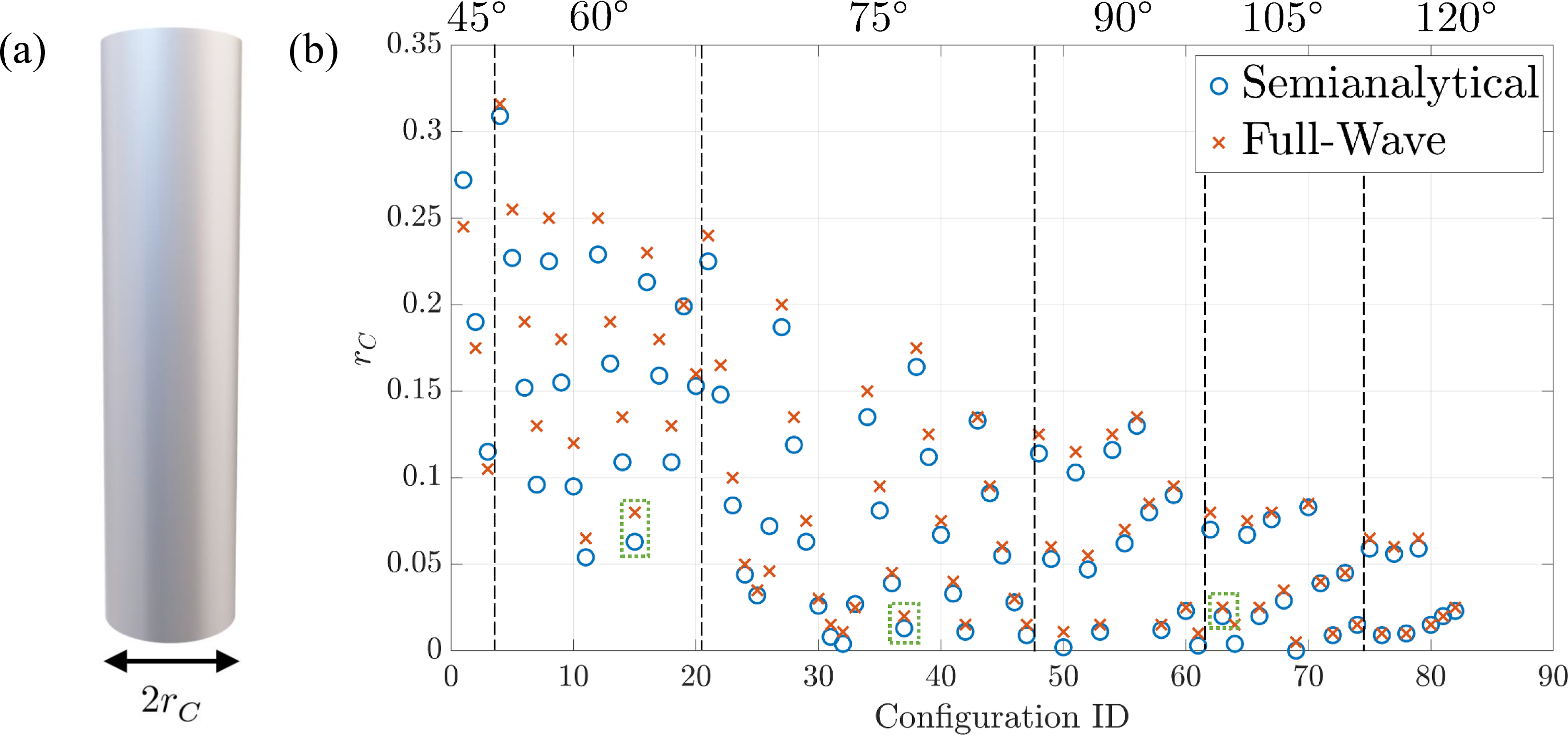}}
\caption{Semianalytical model for a cylindrical post as a scatterer (\cref{subsec:cylinder_radius_model}). (a) Scatterer geometry. (b) Comparison between values of the cylinder radius obtained by the semianalytical model \crefrange{eq22}{eq23} and values for full transmission obtained via full-wave optimization. All semianalytical calculations were performed for $N=8$.}
\label{fig6}
\end{figure}

Clearly, since our analytical method does not assume a predefined scatterer geometry, multiple forms of passive elements can be used as meta-atoms to test for fulfillment of the design requirements\footnote{\textcolor{black}{It should be noted that also partial-height scatterers, i.e., scatterers which are not attached to both bottom and top WG walls, can be used in this regard, and have been verified to yield satisfactory reflection mitigation results across the operational bandwidth (not shown).}}. To demonstrate the versatility of the synthesis method in this sense, in place of the capacitively-loaded wire discussed in the previous subsection, in this subsection we examine the use of a \emph{cylindrical metallic post} (an element often embedded in rectangular WGs to manipulate guided modes \cite{scatterers7,scatterers2,scatterers10}) of radius $r_C$, as illustrated in \cref{fig6}(a).

For convenience, we focus on symmetric WG bends (${a_1 = a_2}$); as discussed in Subsection \ref{subsec:branches}, in these configurations the axis of symmetry is always a solution branch [given that the scatterer location follows the restriction in \cref{eq9}]. Therefore, utilizing our previous observations, we concentrate on placing the scatterer at a distance $r_0$ from the junction corner $O$ on the axis of symmetry, thus removing the need to invoke the full PTL identification procudure, and saving design time. Subsequently, for a given $r_0$, the cylinder radius $r_C$ should be set so that the desired secondary fields are excited and perfect transmission is achieved.

Our proposed semianalytical model for $r_C$ is based on boundary condition considerations. The desired field distribution in the junction, occurring when the reflection vanishes and the transmission satisfies power balance [\crefrange{eq18a}{eq18b}], is calculated semianalytically for the case of an infinitesimal line source placed at a certain PTL and carrying the current $I_\mathrm{NR}$ [\crefrange{eq6}{eq7},\cref{eq11},\cref{eq19}], as discussed in Section \ref{sec:theory}. Since we assume subwavelength scatterers, such that the discrepancy between this theoretical field distribution and the actual field distribution with a passive scatterer is small (as also verified for the cases presented in Figs. 4 and 5), the boundary conditions posed by a potentially suitable metallic post configuration should be compatible with the semianalytically obtained field pattern. In other words, it would be reasonable to choose a value $\tilde{r}_C$ for the radius of this post such that the tangential electric field predicted by \cref{eq6} and \cref{eq7} vanishes on the cylindrical shell, namely, satisfying
\begin{equation}
    \mathbf{E_j}(r_0 + \tilde{r}_C,\phi_0) = 0, \label{eq22}
\end{equation}
where $\tilde{r}_C > 0$. Accordingly, for a given bend configuration, we calculate $\mathbf{E_j}(r,\phi_0)$ in the junction via \crefrange{eq6}{eq7}, with the chosen PTL used to evaluate \cref{eq19}, and identify the values $\tilde{r}_C$ that satisfy \cref{eq22} as potential radii for a cylindrical post meta-atom.

To test this approximate\footnote{Clearly, the simplicity of this model, which stems from taking a single point around $(r_0,\phi_0)$ and checking for vanishing of the electric field, rather than using a more global approach to enforce the boundary conditions, yields a somewhat coarse approximation for the required cylinder radius.} approach, we calculated $\tilde{r}_C$ as described for a collection of WG bend configurations ($N=8$); in parallel, we used ANSYS HFSS with the same configurations to find the \emph{actual} cylinder radius, $r_C$, which provides optimal transmission. The configurations examined were all right-angle bends, featuring WG widths $a=a_1=a_2$ varying from $0.55\lambda$ to $0.95\lambda$, in steps of $0.025\lambda$ (total of 17 configurations). In all configurations, we chose $r_0=a/2$ as the center of the cylinder.

Even though the model presented by \cref{eq22} is very simplistic, the results for these 17 configurations revealed what appears to be a linear relation between the actual optimal radius $r_C$ and the semianalytically calculated one $\tilde{r}_C$, such that this connection can be approximated by the equation
\begin{equation}
    r_C \approx 1.0852\tilde{r}_C. \label{eq23}
\end{equation}

We tested \cref{eq23} for various symmetric bend configurations of bending angles between $45^{\circ}$ and $120^{\circ}$ and multiple WG widths. If \eqref{eq22} yielded more than one value for $\tilde{r}_C$, the smallest value was chosen. Figure 6(b) presents a comparison between the post radius predicted semianalytically via \eqref{eq22} and \cref{eq23} (blue $\circ$), and the optimal geometry deduced from full-wave simulation sweeps (red $\times$). The various configurations (82 in total) have been assigned identification numbers (configuration ID), and the configurations used for obtaining \cref{eq23} were excluded from the figure; the full details on each bend configuration and the quantitative description of the data presented in \cref{fig6}(b) can be found in \cref{sec:app_A}. We observe good agreement between the cylinder radii obtained via the semianalytical method and the full-wave simulations for bending angles of $75^{\circ}$ and above. As the bending angle decreases, however, the semianalytical estimate for $r_C$ becomes less accurate; nonetheless, in most of the $60^{\circ}$ bend configurations, the radius, though deviating slightly more from the optimal radius found in HFSS, provides almost full transmission across the bend (see \cref{table1} in \cref{sec:app_A}). In addition, even when the estimated radius does not yield a satisfactory value for the transmission coefficient, it does provide a good approximation for $r_C$, serving as a favourable starting point around which a brief full-wave sweep can be performed. The discrepancy observed in \cref{fig6}(b) for acute angles may be associated with the relatively large values of $r_C$ required to mitigate reflection loss in these junctions [\cref{fig6}(b)], for which modelling the scatterer as an infinitesimally thin current-carrying wire (Section \ref{sec:theory}), becomes relatively less accurate. All in all, we can conclude that the simplistic semianalytical model proposed herein provides a good estimation method for the post radius, performing well for a wide variety of relevant bend configurations, including different WG widths and bend angles.

\begin{figure}[!t]
\centerline{\includegraphics[width=\columnwidth]{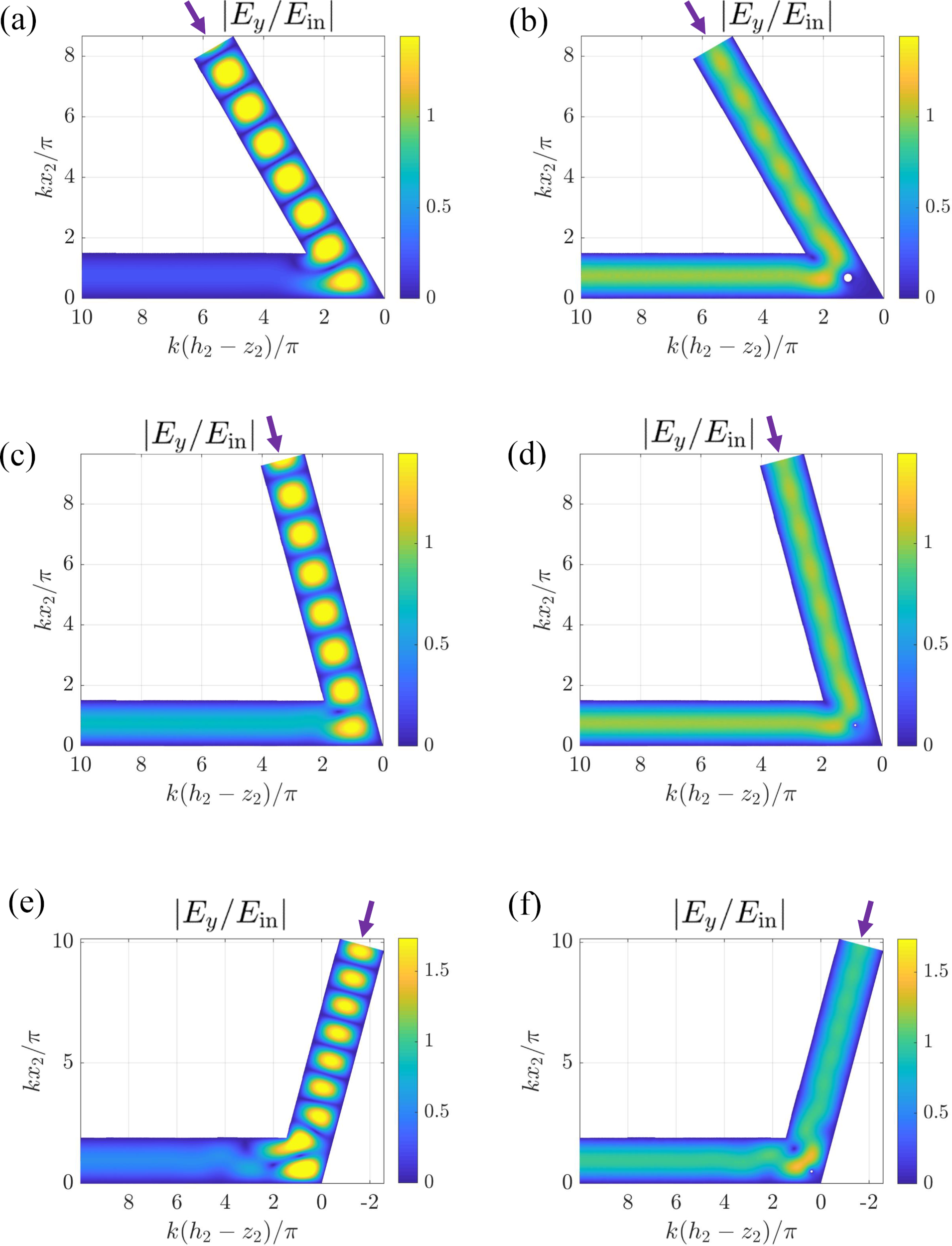}}
\caption{Mitigating reflection loss in WG bends using cylindrical posts [\cref{subsec:cylinder_radius_model}, see geometry in \cref{fig6}(a)]. (a),(c),(e) Full-wave simulated field magnitudes $\left|E_y/E_\mathrm{in}\right|$ for the bare junctions and (b),(d),(f) matched junctions using the semianalytically designed [\crefrange{eq22}{eq23}] cylindrical post for configurations \#15, \#37 and \#63 from \cref{fig6}(b), respectively. In all field plots, the purple arrow indicates the input port. In (b),(d),(f), the white circular void represents the post. The configuration parameters are: \#15, $\Phi=60^{\circ}$, $a=0.75\lambda$, $r_0=11a/12$; \#37, $\Phi=75^{\circ}$, $a=0.75\lambda$, $r_0=3a/4$; \#63, $\Phi=105^{\circ}$, $a=0.95\lambda$, $r_0=a/3$.}
\label{fig7}
\end{figure}

\cref{fig7} showcases three of the tested configurations - \#15, \#37, \#63 [marked by green dotted frames in \cref{fig6}(b)], where the bend angles are $60^{\circ}$, $75^{\circ}$ and $105^{\circ}$, and the reflection of the bare junctions exceeds ${90}\%$, ${50}\%$ and ${70}\%$, respectively. For each case, the total fields obtained from full-wave simulations in the bare junction are presented, along with the simulated fields corresponding to the WG bend configuration with the embedded cylindrical post of the prescribed radius $r_C$ obtained from the semianalytical model [blue $\circ$ in \cref{fig6}(b)] without any further optimization. The WG widths, cylinder location and cylinder radius for all three cases are given in the caption of \cref{fig7}. As can be clearly seen in \cref{fig7}(b), (d), and (f), the improvement in the transmission across the bend is substantial, with the fraction of transmitted power reaching near-unity values after the introduction of the cylindrical post for all three cases. The slight interference pattern in the input port observed mainly in \cref{fig6}(b) and (d) results from very small residual reflected power (less than $1\%$ reflection in all cases, as seen from \cref{table1}).

These examples demonstrate the efficacy of the semianalytical model presented herein, enabling synthesis of seamless symmetric WG bends of various angles and port widths while diminishing the need for long full-wave optimizations, thus offering detailed design solutions featuring a standard simple cylindrical post with specified radius and location in the junction.

\subsection{Practical Considerations}
\label{subsec:practical}

\textcolor{black}{As the design scheme is concluded, it is important to discuss practical considerations related to the performance of the modified junction solutions designed via the proposed methods. In particular, we consider two important aspects, addressing the device bandwidth (frequency response) and the sensitivity to fabrication tolerances (possible deviations in scatterer location).}

\begin{figure}[!b]
\centerline{\includegraphics[width=\columnwidth]{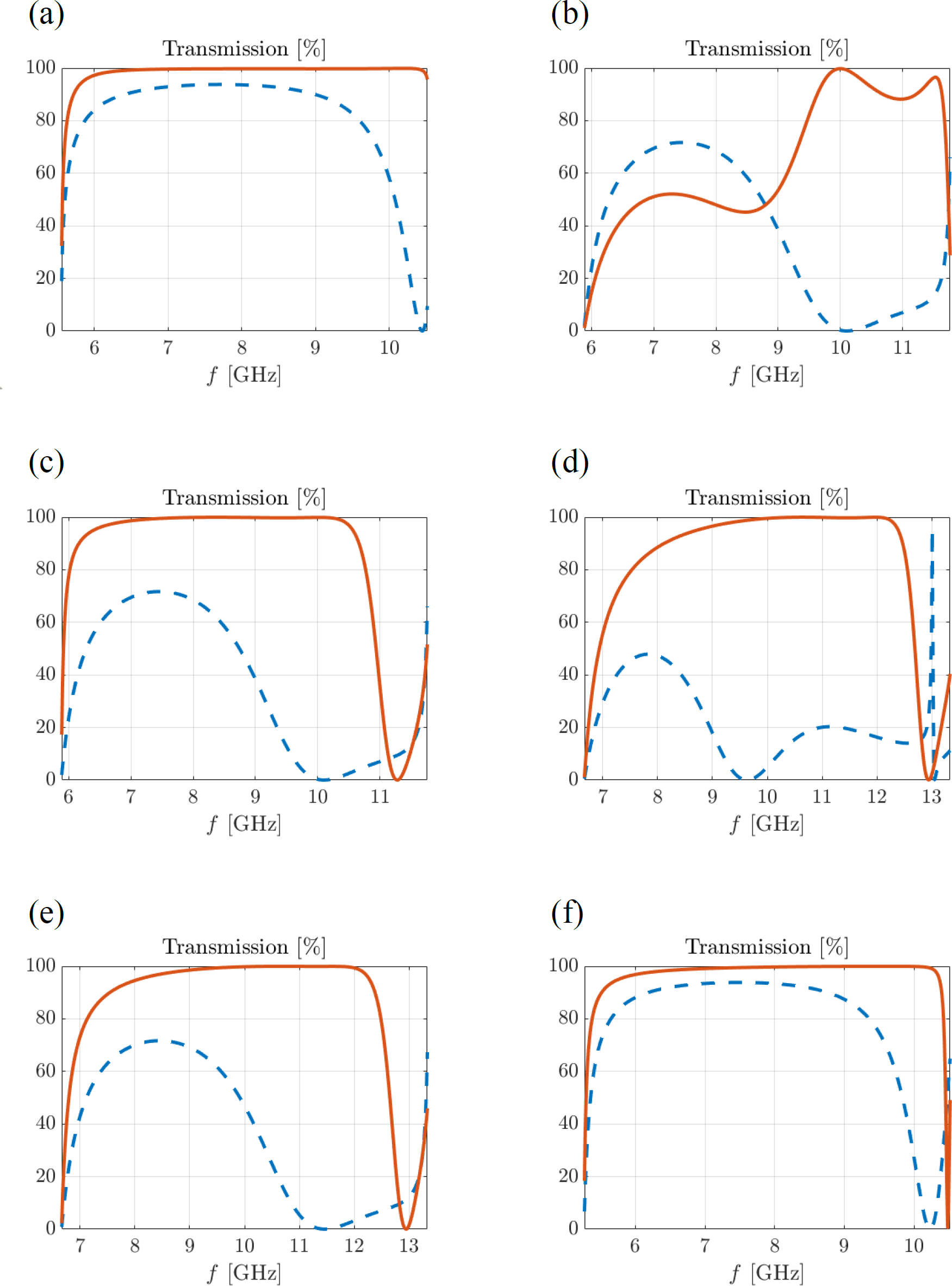}}
\caption{\textcolor{black}{Bandwidth performance of the modified junction solutions presented throughout the paper. Each graph includes the transmission of both the bare junction (dashed blue line) and the junction with the proposed scatterer (solid orange line), covering the entire single-mode operation frequency interval. The frequency responses correspond to the case studies presented in (a) Fig. 4; (b) Fig. 5(c); (c) Fig. 5(e); (d) Fig, 7(b); (e) Fig. 7(d); (f) Fig. 7(f), considering implementations with both (a)-(c) loaded wire and (d)-(f) cylindrical posts.}}
\label{fig8}
\end{figure}

\textcolor{black}{\cref{fig8} presents the frequency response of all the configurations whose electric field distribution is showcased in Figs. 4(d), 5(c), 5(e), 7(b), 7(d) and 7(f), covering the entire single-mode operation bandwidth for each case, from cutoff to the appearance of the second-order mode. For reference, the frequency response of the corresponding bare junctions is presented as well for each case. As can be seen, in all configurations where the scatterer was placed on the axis of symmetry (or, as in Fig 4, at a distorted “generalized” axis of symmetry), we encounter transmission of $90\%$ or above in almost all the relevant frequency band. When the scatterer is placed at the “opposing” solution branch [\cref{fig8}(b)], the bandwidth is reduced; however, even in this case, the solution still provides dramatically enhanced transmission with respect to the bare junction (${\left|S_{21}\right|}^2 \geq 88\%$ in a $~2\mathrm{GHz}$ range around $f=10.6\mathrm{GHz}$). It should be noted that in all cases we did not employ any special means to target bandwidth optimization. The resulting relatively broadband response is typical to other previously reported MG-based solutions, which are relying on a small number of non-resonant elements \cite{PhysRevLett.119.067404,Rabinovich_2}.}

\begin{figure}[!b]
\centerline{\includegraphics[width=\columnwidth]{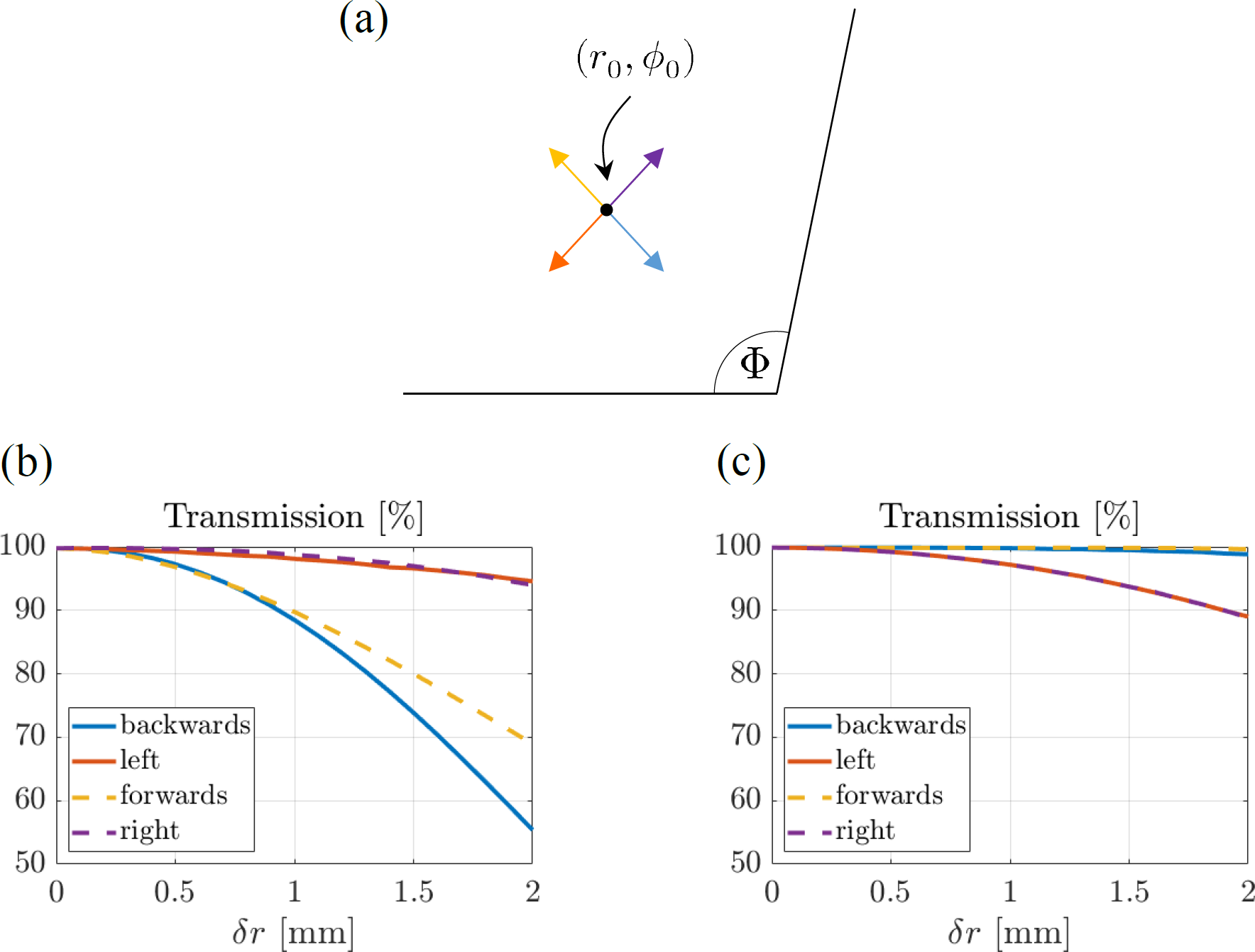}}
\caption{\textcolor{black}{Sensitivity to fabrication errors in scatterer location. (a) Illustration of tested locations, in four directions away from the PTL; (b),(c) Sensitivity to shifts in the scatterer position away from the PTL in each of the directions, dictated by the color matching the arrow color in (a), for the second case study of the $75^{\circ}$ bend, with each of the PTL choices, presented in Figs. 5(c) and 5(e), respectively.}}
\label{fig9}
\end{figure}

\textcolor{black}{To examine the system robustness to the scatterer location, we considered the two modified junction configurations showcased in \cref{fig5}(c), and 5(e); these serve as good representative case studies, since they consider PTLs off and on the axis of symmetry, respectively. For each configuration, we evaluate the device performance when the scatterer is shifted by a distance $\delta r$ from the chosen PTL. We tested four possible deviation paths along two dimensions around the PTL [\cref{fig9}(a)]: two of them correspond to shifts in the radial coordinate, i.e. $(r_0+\delta r,\phi_0)$, $(r_0-\delta r,\phi_0)$, and the other two correspond to changes along the orthogonal direction, namely $(\sqrt{r_0^2+\delta r^2},\phi_0 + \tan^{-1}{(\delta r / r_0)} )$, $(\sqrt{r_0^2+\delta r^2},\phi_0 - \tan^{-1}{(\delta r / r_0)} )$. Along each one of these paths, we calculated the system transmission for $\delta r$ varying between $0.1\mathrm{mm}\approx \lambda/300$ and $2\mathrm{mm}\approx \lambda/15$.}

\textcolor{black}{First, we performed this test on the configuration whose electric field distribution is presented in \cref{fig5}(c). We moved the \emph{exact same} scatterer we used there around the chosen PTL, where the shift along the radial direction [forward/backward in \cref{fig9}(a)] can be seen as moving the scatterer \emph{away} from the solution branch, and the change in the orthogonal (~azimuthal) direction [left/right in \cref{fig9}(a)] can be seen as moving the scatterer \emph{within} the solution branch. The results are presented in \cref{fig9}(b). As can be seen, while the performance deteriorates more when the spatial deviation drives the scatterer outside the solution branch, for fabrication errors of up to $1\mathrm{mm}$ (well within typical CNC tolerances \cite{velling_2021}), the transmission across the bend remains very high for any of the considered discrepancies. For the case study of \cref{fig5}(e), where the original PTL is located on the symmetry axis, the resilience to such errors is even better, as demonstrated in \cref{fig9}(c). For both deviation paths, within the solution branch [radially in this case, forward/backward in \cref{fig9}(a)] or away from it (left/right), ${\left|S_{21}\right|}^2 \geq 90\%$ even for $\delta r \approx 2\mathrm{mm}$. In both examined cases [\cref{fig9}(b) and (c)], we observe that scatterer deviations away from the solution branch affect more the solution’s efficacy than deviations within the solution branch, as expected.}

\textcolor{black}{Considering other reports using common methods (i.e., simple scatterers or bend deformations) to mitigate reflections across H-plane waveguide bends \cite{Casanueva2004,Ma1997,Alimenti1994,Coccioli1996,partialheightcompensate}, the performance exhibited by our solution when applied to the same waveguide configurations is mostly comparable in terms of the $-20 \mathrm{dB}$ reflection bandwidth (where ${\left|S_{11}\right|}^2<1\%$). In applications where a more demanding reflection suppression is required, introducing additional degrees of freedom to our methodology, as has been done previously in free-space metagratings \cite{Rabinovich_3}, is expected to enable further enhancement of the transmission efficiency.}

\textcolor{black}{Overall, these results indicate that the proposed solution is practically viable, demonstrating both robustness to fabrication tolerances and wide operational bandwidth. From the explored case studies, it may be concluded that positioning the scatterers along a symmetry axis (see discussion in \cref{subsec:branches}) is preferable, and is expected to lead to higher performance in these regards.}

\section{Conclusion}
\label{sec:conclusion}
In this paper, we have used the methodology associated with MG analysis and synthesis to propose a simple and flexible approach for the suppression of reflections in abrupt H-plane single-mode rectangular WG bends, of arbitrary bend angles and WG widths. Using modal techniques, we have analyzed the response of the bend to a $\mathrm{TE}_{10}$ excitation, considering also the contribution of a current line source embedded in the junction. Subsequently, we have found potential locations for the line source, in which the fields produced by it yield full transmission across the bend while maintaining the power balance in the system; such a current source may therefore be replaced by a passive polarizable element that implements the same functionality.

The semianalytical methodology was verified using full-wave simulations, demonstrating that, indeed, the desired mitigation of reflection loss can be efficiently achieved with either impressed current sources or (more realistic) passive polarizable elements (realized as capacitively-loaded wires or cylindrical metallic posts), the location of which is accurately found via the model. The model was further harnessed for providing physical insight regarding the nature of the solution branches associated with potential scatterer location, yielding guidelines for identifying favourable embedding spots for practical design with finite conductor losses.

Finally, we have proposed a semianalytical method to estimate the detailed scatterer geometry (cylindrical post radius) required to eliminate reflection loss in symmetric bend configuration, shown to be applicable for a broad range of bend angles and WG widths. In these cases, similarly to previously reported MGs, a complete design of the configuration, including the scatterer location and its dimensions, can be obtained using the semianalytical model, avoiding extensive full-wave optimizations. \textcolor{black}{We stress that, in contrast to previous reports, our method is of general nature, accommodating a wide range of bend angles, scatterer geometries, and input/output asymmetry; in this regard, it yields a simple solution in the form of a single, purely-reactive, small vertical inclusion, guaranteed \emph{rigorously}, and requires no deformation of the junction layout whatsoever. The proposed design also exhibits wideband characteristics and shows robustness to fabrication inaccuracies.} In addition to introducing an appealing semianalytical tool for devising a simple and effective solution to the WG bend problem, these results, verified using commercial general-purpose solvers, highlight the efficacy and usefulness of the MG synthesis approach, allowing tackling diverse electromagnetic problems beyond beam manipulation in free-space.

\captionsetup[supertabular]{labelfont=sc,textfont=sc,labelsep=newline,justification=centering}

\onecolumn
\begin{center}
\tablefirsthead{%
 \hline\hline
 $\Phi [^{\circ}]$ & $a [\lambda]$ & $T_B [\%]$ & \textbf{ID} & $r_0/a$ & $r_C [\lambda]$ (semianalytical) & $T_M [\%]$ & $r_C [\lambda]$ (Full-Wave) & $T_H [\%]$ \\
 \hline\hline }
\tablehead{%
\hline
\multicolumn{9}{|l|}{\small\sl continued from previous page}\\ \hline
\hline
 $\Phi [^{\circ}]$ & $a [\lambda]$ & $T_B [\%]$ & \textbf{ID} & $r_0/a$ & $r_C [\lambda]$ (semianalytical) & $T_M [\%]$ & $r_C [\lambda]$ (Full-Wave) & $T_H [\%]$ \\
 \hline\hline }
\tabletail{%
\hline\hline
\multicolumn{9}{|r|}{\small\sl continued on next page}\\
\hline}
\tablelasttail{\hline}
\tablecaption{Bend Configurations Used to Test the semianalytical Model \protect\linebreak for the Cylinder Radius, Corresponding to \cref{fig6}(b) \vspace{0.3cm}}
\label{table1}
\begin{supertabular}{|c|c|c|c|c|c|c|c|c| }

\multirow{3}{*}{\textcolor{black}{$45$}}&\multirow{3}{*}{\textcolor{black}{$0.95$}}&\multirow{3}{*}{\textcolor{black}{$42.1$}}&\textcolor{black}{$\mathbf{1}$}&\textcolor{black}{$9/12$}&\textcolor{black}{$0.272$}&\textcolor{black}{$0$}&\textcolor{black}{$0.245$}&\textcolor{black}{$99.61$}\\ \cline{4-9}
&&&\textcolor{black}{$\mathbf{2}$}&\textcolor{black}{$10/12$}&\textcolor{black}{$0.19$}&\textcolor{black}{$17$}&\textcolor{black}{$0.175$}&\textcolor{black}{$99.45$}\\ \cline{4-9}
&&&\textcolor{black}{$\mathbf{3}$}&\textcolor{black}{$11/12$}&\textcolor{black}{$0.115$}&\textcolor{black}{$63.56$}&\textcolor{black}{$0.105$}&\textcolor{black}{$99.68$}\\ \hline
\multirow{4}{*}{$60$}&\multirow{4}{*}{$0.95$}&\multirow{4}{*}{$14.19$}&$\mathbf{4}$&$8/12$&$0.309$&$92.71$&$0.316$&$96.91$\\ \cline{4-9}
&&&$\mathbf{5}$&$9/12$&$0.227$&$82.85$&$0.255$&$99.85$\\ \cline{4-9}
&&&$\mathbf{6}$&$10/12$&$0.152$&$73.01$&$0.19$&$99.94$\\ \cline{4-9}
&&&$\mathbf{7}$&$11/12$&$0.096$&$70.98$&$0.13$&$99.86$\\ \hline
\multirow{4}{*}{$60$}&\multirow{4}{*}{$0.85$}&\multirow{4}{*}{$19.94$}&$\mathbf{8}$&$8/12$&$0.225$&$98.53$&$0.25$&$99.97$\\ \cline{4-9}
&&&$\mathbf{9}$&$9/12$&$0.155$&$97.58$&$0.18$&$99.97$\\ \cline{4-9}
&&&$\mathbf{10}$&$10/12$&$0.095$&$97.55$&$0.12$&$99.98$\\ \cline{4-9}
&&&$\mathbf{11}$&$11/12$&$0.054$&$98.82$&$0.065$&$99.97$\\ \hline
\multirow{4}{*}{$60$}&\multirow{4}{*}{$0.75$}&\multirow{4}{*}{$4.97$}&$\mathbf{12}$&$8/12$&$0.229$&$99.64$&$0.25$&$99.99$\\ \cline{4-9}
&&&$\mathbf{13}$&$9/12$&$0.166$&$99.44$&$0.19$&$99.99$\\ \cline{4-9}
&&&$\mathbf{14}$&$10/12$&$0.109$&$99.34$&$0.135$&$99.99$\\ \cline{4-9}
&&&$\mathbf{15}$&$11/12$&$0.063$&$99.54$&$0.08$&$99.99$\\ \hline
\multirow{3}{*}{$60$}&\multirow{3}{*}{$0.65$}&\multirow{3}{*}{$32.72$}&$\mathbf{16}$&$9/12$&$0.213$&$99.78$&$0.23$&$99.99$\\ \cline{4-9}
&&&$\mathbf{17}$&$10/12$&$0.159$&$99.67$&$0.18$&$99.99$\\ \cline{4-9}
&&&$\mathbf{18}$&$11/12$&$0.109$&$99.61$&$0.13$&$99.99$\\ \hline
\multirow{2}{*}{$60$}&\multirow{2}{*}{$0.55$}&\multirow{2}{*}{$42.06$}&$\mathbf{19}$&$10/12$&$0.199$&$99.95$&$0.2$&$99.97$\\ \cline{4-9}
&&&$\mathbf{20}$&$11/12$&$0.153$&$99.78$&$0.16$&$99.97$\\ \hline
\multirow{6}{*}{$75$}&\multirow{6}{*}{$0.95$}&\multirow{6}{*}{$8.41$}&$\mathbf{21}$&$5/12$&$0.225$&$98.65$&$0.24$&$99.95$\\ \cline{4-9}
&&&$\mathbf{22}$&$6/12$&$0.148$&$96.76$&$0.165$&$99.96$\\ \cline{4-9}
&&&$\mathbf{23}$&$7/12$&$0.084$&$96.44$&$0.1$&$99.97$\\ \cline{4-9}
&&&$\mathbf{24}$&$8/12$&$0.044$&$98.8$&$0.05$&$99.95$\\ \cline{4-9}
&&&$\mathbf{25}$&$9/12$&$0.032$&$99.71$&$0.035$&$99.38$\\ \cline{4-9}
&&&$\mathbf{26}$&$10/12$&$0.072$&$51.92$&$0.046$&$99.19$\\ \hline
\multirow{7}{*}{$75$}&\multirow{7}{*}{$0.85$}&\multirow{7}{*}{$0.26$}&$\mathbf{27}$&$5/12$&$0.187$&$99.76$&$0.2$&$99.99$\\ \cline{4-9}
&&&$\mathbf{28}$&$6/12$&$0.119$&$99.56$&$0.135$&$99.99$\\ \cline{4-9}
&&&$\mathbf{29}$&$7/12$&$0.063$&$99.63$&$0.075$&$99.99$\\ \cline{4-9}
&&&$\mathbf{30}$&$8/12$&$0.026$&$99.79$&$0.03$&$99.97$\\ \cline{4-9}
&&&$\mathbf{31}$&$9/12$&$0.008$&$97.71$&$0.015$&$99.94$\\ \cline{4-9}
&&&$\mathbf{32}$&$10/12$&$0.004$&$82.48$&$0.011$&$99.57$\\ \cline{4-9}
&&&$\mathbf{33}$&$11/12$&$0.027$&$99$&$0.025$&$99.85$\\ \hline
\multirow{4}{*}{$75$}&\multirow{4}{*}{$0.75$}&\multirow{4}{*}{$46.9$}&$\mathbf{34}$&$6/12$&$0.135$&$99.82$&$0.15$&$99.99$\\ \cline{4-9}
&&&$\mathbf{35}$&$7/12$&$0.081$&$99.81$&$0.095$&$99.99$\\ \cline{4-9}
&&&$\mathbf{36}$&$8/12$&$0.039$&$99.9$&$0.045$&$99.98$\\ \cline{4-9}
&&&$\mathbf{37}$&$9/12$&$0.013$&$99.81$&$0.02$&$99.95$\\ \hline
\multirow{5}{*}{$75$}&\multirow{5}{*}{$0.65$}&\multirow{5}{*}{$71.29$}&$\mathbf{38}$&$6/12$&$0.164$&$99.92$&$0.175$&$99.98$\\ \cline{4-9}
&&&$\mathbf{39}$&$7/12$&$0.112$&$99.86$&$0.125$&$99.99$\\ \cline{4-9}
&&&$\mathbf{40}$&$8/12$&$0.067$&$99.86$&$0.075$&$99.98$\\ \cline{4-9}
&&&$\mathbf{41}$&$9/12$&$0.033$&$99.92$&$0.04$&$99.98$\\ \cline{4-9}
&&&$\mathbf{42}$&$10/12$&$0.011$&$99.79$&$0.015$&$99.98$\\ \hline
\multirow{5}{*}{$75$}&\multirow{5}{*}{$0.55$}&\multirow{5}{*}{$58.57$}&$\mathbf{43}$&$7/12$&$0.133$&$99.94$&$0.135$&$99.98$\\ \cline{4-9}
&&&$\mathbf{44}$&$8/12$&$0.091$&$99.82$&$0.095$&$99.98$\\ \cline{4-9}
&&&$\mathbf{45}$&$9/12$&$0.055$&$99.86$&$0.06$&$99.98$\\ \cline{4-9}
&&&$\mathbf{46}$&$10/12$&$0.028$&$99.92$&$0.03$&$99.97$\\ \cline{4-9}
&&&$\mathbf{47}$&$11/12$&$0.009$&$99.34$&$0.015$&$99.79$\\ \hline
\multirow{3}{*}{$90$}&\multirow{3}{*}{$0.95$}&\multirow{3}{*}{$1.35$}&$\mathbf{48}$&$4/12$&$0.114$&$99.42$&$0.125$&$99.98$\\ \cline{4-9}
&&&$\mathbf{49}$&$5/12$&$0.053$&$99.53$&$0.06$&$99.96$\\ \cline{4-9}
&&&$\mathbf{50}$&$8/12$&$0.002$&$32.32$&$0.011$&$99.47$\\ \hline
\multirow{3}{*}{$90$}&\multirow{3}{*}{$0.85$}&\multirow{3}{*}{$54.28$}&$\mathbf{51}$&$4/12$&$0.103$&$99.89$&$0.115$&$99.99$\\ \cline{4-9}
&&&$\mathbf{52}$&$5/12$&$0.047$&$99.91$&$0.055$&$99.99$\\ \cline{4-9}
&&&$\mathbf{53}$&$10/12$&$0.011$&$97.77$&$0.015$&$99.73$\\ \hline
\multirow{2}{*}{$90$}&\multirow{2}{*}{$0.75$}&\multirow{2}{*}{$82.71$}&$\mathbf{54}$&$4/12$&$0.116$&$99.94$&$0.125$&$99.99$\\ \cline{4-9}
&&&$\mathbf{55}$&$5/12$&$0.062$&$99.93$&$0.07$&$99.99$\\ \hline
\multirow{3}{*}{$90$}&\multirow{3}{*}{$0.65$}&\multirow{3}{*}{$85.71$}&$\mathbf{56}$&$4/12$&$0.13$&$99.97$&$0.135$&$99.99$\\ \cline{4-9}
&&&$\mathbf{57}$&$5/12$&$0.08$&$99.94$&$0.085$&$99.99$\\ \cline{4-9}
&&&$\mathbf{58}$&$7/12$&$0.012$&$99.95$&$0.015$&$99.98$\\ \hline
\multirow{3}{*}{$90$}&\multirow{3}{*}{$0.55$}&\multirow{3}{*}{$72.69$}&$\mathbf{59}$&$5/12$&$0.09$&$99.95$&$0.095$&$99.96$\\ \cline{4-9}
&&&$\mathbf{60}$&$7/12$&$0.023$&$99.94$&$0.025$&$99.98$\\ \cline{4-9}
&&&$\mathbf{61}$&$8/12$&$0.003$&$98.47$&$0.01$&$99.74$\\ \hline
\multirow{3}{*}{$105$}&\multirow{3}{*}{$0.95$}&\multirow{3}{*}{$27.59$}&$\mathbf{62}$&$3/12$&$0.07$&$99.85$&$0.08$&$99.98$\\ \cline{4-9}
&&&$\mathbf{63}$&$4/12$&$0.02$&$99.92$&$0.025$&$99.97$\\ \cline{4-9}
&&&$\mathbf{64}$&$8/12$&$0.004$&$81.59$&$0.015$&$99.79$\\ \hline
\multirow{2}{*}{$105$}&\multirow{2}{*}{$0.85$}&\multirow{2}{*}{$88.16$}&$\mathbf{65}$&$3/12$&$0.067$&$99.95$&$0.075$&$99.99$\\ \cline{4-9}
&&&$\mathbf{66}$&$4/12$&$0.02$&$99.98$&$0.025$&$99.99$\\ \hline
\multirow{3}{*}{$105$}&\multirow{3}{*}{$0.75$}&\multirow{3}{*}{$93.55$}&$\mathbf{67}$&$3/12$&$0.076$&$99.97$&$0.08$&$99.99$\\ \cline{4-9}
&&&$\mathbf{68}$&$4/12$&$0.029$&$99.98$&$0.035$&$99.99$\\ \cline{4-9}
&&&$\mathbf{69}$&$5/12$&$0.00005$&$97.51$&$0.005$&$99.97$\\ \hline
\multirow{3}{*}{$105$}&\multirow{3}{*}{$0.65$}&\multirow{3}{*}{$93.15$}&$\mathbf{70}$&$3/12$&$0.083$&$99.98$&$0.085$&$99.99$\\ \cline{4-9}
&&&$\mathbf{71}$&$4/12$&$0.039$&$99.98$&$0.04$&$99.98$\\ \cline{4-9}
&&&$\mathbf{72}$&$5/12$&$0.009$&$99.96$&$0.01$&$99.98$\\ \hline
\multirow{2}{*}{$105$}&\multirow{2}{*}{$0.55$}&\multirow{2}{*}{$84.57$}&$\mathbf{73}$&$4/12$&$0.045$&$99.96$&$0.045$&$99.96$\\ \cline{4-9}
&&&$\mathbf{74}$&$5/12$&$0.015$&$99.94$&$0.015$&$99.94$\\ \hline
\multirow{2}{*}{$120$}&\multirow{2}{*}{$0.95$}&\multirow{2}{*}{$84.57$}&$\mathbf{75}$&$2/12$&$0.059$&$99.97$&$0.065$&$99.99$\\ \cline{4-9}
&&&$\mathbf{76}$&$3/12$&$0.009$&$99.94$&$0.01$&$99.96$\\ \hline
\multirow{2}{*}{$120$}&\multirow{2}{*}{$0.85$}&\multirow{2}{*}{$96.74$}&$\mathbf{77}$&$2/12$&$0.056$&$99.98$&$0.06$&$99.99$\\ \cline{4-9}
&&&$\mathbf{78}$&$3/12$&$0.01$&$99.98$&$0.01$&$99.98$\\ \hline
\multirow{2}{*}{$120$}&\multirow{2}{*}{$0.75$}&\multirow{2}{*}{$97.66$}&$\mathbf{79}$&$2/12$&$0.059$&$99.99$&$0.065$&$99.99$\\ \cline{4-9}
&&&$\mathbf{80}$&$3/12$&$0.015$&$99.99$&$0.015$&$99.99$\\ \hline
$120$&$0.65$&$97.15$&$\mathbf{81}$&$3/12$&$0.02$&$99.99$&$0.02$&$99.99$\\ \hline
$120$&$0.55$&$92.93$&$\mathbf{82}$&$3/12$&$0.023$&$99.98$&$0.025$&$99.97$\\ \hline\hline

\end{supertabular}
\end{center}
\twocolumn

\appendices
\crefalias{section}{appendix}
\section{Verification of the Simplistic Model for a Cylindrical Post Meta-Atom}
\label{sec:app_A}

In Subsection \ref{subsec:cylinder_radius_model} we proposed a simplistic model, allowing estimation of the radius of a cylindrical post to be used as a scatterer to facilitate perfect transmission across the bend. This Appendix provides further details regarding the steps taken in order to evaluate the range of validity of this model.

As discussed in Subsection \ref{subsec:cylinder_radius_model}, we considered 82 different configurations of bends with scatterers to test the proposed estimation procedure; the comparison between the semianalytically estimated post radii and the actual optimal ones (from full-wave simulations) graphically appears in \cref{fig6}(b). The complete list of configuration, including all quantitative data, is included in Table I herein.

The configuration list includes bending angles ranging from $45^{\circ}$ to $120^{\circ}$ with a step size of $15^{\circ}$; for each bending angle, the WG width $a=a_1=a_2$ is ranging from $0.55\lambda$ to $0.95\lambda$ with a step size of $0.1\lambda$ (total of 30 different scattering WG junctions to match)\footnote{As in all the cases discussed in this paper, the WG height in these configurations was $b=0.25\lambda$, and the operating frequency was $f=10\mathrm{GHz}$.} For each of these bend parameter combinations, the model was tested for multiple distances $r_0$ of the cylinder center from the bend corner (where the cylinder was placed at the bend's axis of symmetry), such that $r_0$ was ranging from $a/12$ to $11a/12$ with a typical step size of $a/12$. For the purpose of our synthesis procedure, a calculated value for $r_C$ via \cref{eq22} and \cref{eq23} is considered valid only if: $\tilde{r}_C$ retrieved from \cref{eq22} is positive; $r_0$ satisfies \cref{eq9}\footnote{This condition for validity of $r_0$ is crucial, since as $r_0$ increases going up to $11a/12$, it can become too large such that it exceeds the value of $h_1(=h_2)$.}; and $r_0 \sin(\Phi/2) > r_C$, namely the cylinder fits inside the WG. When multiple valid values for $r_C$ were found from the semianalytical model, the smallest one was used for the tests. In addition, as indicated in Subsection \ref{subsec:cylinder_radius_model}, right-angle bends with $r_0=6a/12$ are not included in the table since they were used for the construction of the semianalytical model yielding \cref{eq23}.

For each configuration considered in this testing, \cref{table1} details the transmission across the bend before and after the inclusion of the designed scatterer\footnote{All posts in these tests are made of copper.}; for the latter case, the performance of a cylindrical post with the semianalytically estimated radius "$r_C$ (semianalytical)" is compared to the optimal one "$r_C$ (Full-Wave)" obtained via full-wave sweeps. In the presented results in \cref{table1}, the fraction of power transmitted to the output in the bare junction is denoted by $T_B$. The transmission for the radius values $r_C$ (semianalytical) and $r_C$ (Full-Wave) are denoted by $T_M$ and $T_H$, respectively. The overall results show that for bends of $75^{\circ}$ or more, the semianalytical model yields a very good estimation for the optimal cylinder radius $r_C$. For smaller bend angles, even though $T_M$ is low in some configurations, the semianalytically obtained value provides a good starting point for a parametric sweep in a full-wave solver. Moreover, out of the 30 scattering WG junctions described above, a design solution was produced for almost every junction, and in the vast majority of cases it even enabled design flexibility by providing multiple choices of meta-atom locations and radii.

\bibliographystyle{IEEEtran}

\end{document}